\documentclass[10pt,twocolumn,letterpaper]{article}
\usepackage[pagenumbers]{cvpr} %

\usepackage{graphicx}
\usepackage{amsmath}
\usepackage{amssymb}
\usepackage{booktabs}
\usepackage{comment}
\usepackage{multirow}
\usepackage{tabularx}
\usepackage{float}
\usepackage{xcolor}
\usepackage{bbm}
\usepackage{import}
\usepackage{tikz}
\usepackage{comment}

\usepackage{pifont} %
\usepackage{booktabs}
\usepackage{multirow}
\usepackage{adjustbox}
\usepackage{fontawesome} %

\definecolor{opA}{rgb}{0.9,0.6,0.0}
\definecolor{opB}{rgb}{0.35,0.70,0.90}
\definecolor{opC}{rgb}{0.8,0.40,0.0}
\definecolor{opD}{rgb}{0.0,0.60,0.50} %
\definecolor{opE}{rgb}{0.8,0.6,0.7}
\definecolor{opF}{rgb}{0.,0.45,0.70} 

\definecolor{pltBlue}{rgb}{0.12156862745098039, 0.4666666666666667, 0.7058823529411765}
\definecolor{pltOrange}{rgb}{1.0, 0.4980392156862745, 0.054901960784313725}
\definecolor{pltGreen}{rgb}{0.17254901960784313, 0.6274509803921569, 0.17254901960784313}
\definecolor{pltRed}{rgb}{0.8392156862745098, 0.15294117647058825, 0.1568627450980392}
\definecolor{pltViolet}{rgb}{0.5803921568627451, 0.403921568627451, 0.7411764705882353}
\definecolor{pltBrown}{rgb}{0.5490196078431373, 0.33725490196078434, 0.29411764705882354}
\definecolor{pltMagenta}{rgb}{0.8901960784313725, 0.4666666666666667, 0.7607843137254902}
\definecolor{pltGray}{rgb}{0.4980392156862745, 0.4980392156862745, 0.4980392156862745}
\definecolor{pltLightGreen}{rgb}{0.7372549019607844, 0.7411764705882353, 0.13333333333333333}
\definecolor{pltCyan}{rgb}{0.09019607843137255, 0.7450980392156863, 0.8117647058823529}
\definecolor{pltPink}{rgb}{0.49803921568627, 0.49803921568627, 0.49803921568627}

\definecolor{cmarkcolor}{rgb}{0.49,0.74,0.49}
\definecolor{xmarkcolor}{rgb}{0.86,0.34,0.34}

\newcommand{\xmark}{\textcolor{xmarkcolor}{\ding{55}}}

\def\link#1{
    \ifx&#1&
        \xmark{}
    \else
        {\href{#1}{\faExternalLink}}
    \fi
}

\usepackage{siunitx}
\usepackage{mathrsfs}
\usepackage{subcaption}
\usepackage{pgfplots}
\usepackage{layouts}
\usepackage{algorithm}
\usepackage{algpseudocode}
\usepackage{amsmath}

\definecolor{cvprblue}{rgb}{0.21,0.49,0.74}
\usepackage[pagebackref,breaklinks,colorlinks,citecolor=cvprblue]{hyperref}

\title{3DiFACE: Synthesizing and Editing Holistic 3D Facial Animation} 

\author{
    \begin{minipage}{\textwidth}
        \centering
        Balamurugan Thambiraja\textsuperscript{1,3} \hfill 
        Malte Prinzler\textsuperscript{1,2,4} \hfill  
        Sadegh Aliakbarian\textsuperscript{5} \hfill 
        Darren Cosker\textsuperscript{5} \hfill  
        Justus Thies\textsuperscript{1,3} \\
    \end{minipage}
     \\
    \textsuperscript{1}Max Planck Institute for Intelligent Systems \quad
    \textsuperscript{2}Max Planck ETH Center for Intelligent Systems  \\
        \textsuperscript{3}Technical University of Darmstadt \quad 
        \textsuperscript{4}ETH Zürich \quad  
        \textsuperscript{5}Microsoft Mixed Reality \& AI Lab, UK \\
}

\begin{document}

\twocolumn[{%
\renewcommand\twocolumn[1][]{#1}%
\maketitle
\begin{center}
    \centering
    \captionsetup{type=figure}
    \vspace{-0.6cm}
    \includegraphics[width=0.95\textwidth]{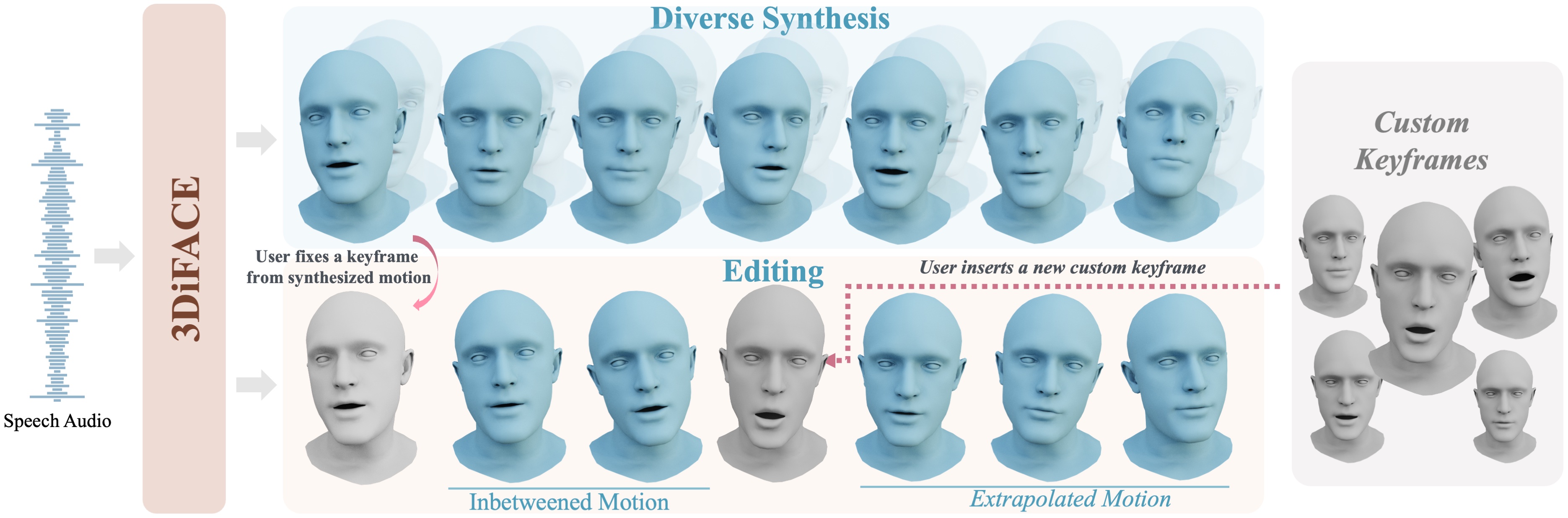}
      \caption{
        \textit{3DiFACE} is a novel diffusion-based method for synthesizing holistic 3D facial animation from an audio input (top). In addition, users can seamlessly edit a synthesized or existing facial animation by defining part of the input as keyframes or by inserting new custom keyframes. These custom keyframes can be  either manually created or sourced from an existing motion database. %
        }
      \vspace{-0.1cm}
      \label{fig:teaser}
\end{center}%
 }]

\begin{abstract}
\vspace{-0.3cm}
Creating personalized 3D animations with precise control and realistic head motions remains challenging for current speech-driven 3D facial animation methods.
\emph{Editing} these animations is especially complex and time consuming, requires precise control and typically handled by highly skilled animators.
Most existing works focus on controlling style or emotion of the synthesized animation and cannot edit/regenerate parts of an input animation. 
They also overlook the fact that multiple plausible lip and head movements can match the same audio input.
To address these challenges, we present 3DiFACE, a novel method for holistic speech-driven 3D facial animation. 
Our approach produces \emph{diverse plausible lip and head motions} for a single audio input and  allows for \emph{editing via keyframing and interpolation}.
Specifically, we propose a fully-convolutional diffusion model that can leverage the viseme-level diversity in our training corpus.
Additionally, we employ a speaking-style personalization and a novel sparsely-guided motion diffusion to enable precise control and editing.
Through quantitative and qualitative evaluations, we demonstrate that our method is capable of generating and editing diverse holistic 3D facial animations given a single audio input, with control between high fidelity and diversity. Code and models are available here:  \href{https://balamuruganthambiraja.github.io/3DiFACE}{https://balamuruganthambiraja.github.io/3DiFACE}
\end{abstract}

\vspace{-0.5cm}
\section{Introduction}\label{sec:intro}
Holistic 3D facial animation transforms digital figures into expressive characters, pivotal for compelling narratives in films and games.
Artists craft these animations with great detail, using precise and iterative editing to ensure every glance and nod adds to the narrative.
In this context, `holistic' refers to 3D facial animation that includes both lip and head movements.
Early works~\cite{cohen01_avsp, edwards2016jali} used procedural-rule based systems to map audio features with facial animation parameters, giving artists precise control and the ability to edit specific-parts of the animation sequence.
However, this process is manual and labour intensive.

With advancements in machine learning, new learning-based methods have emerged that allow for quicker audio-driven facial animations~\cite{zhang20213d, peng2023emotalk, voca, meshtalk, fan2022faceformer, xing2023codetalker}.
However, these methods mainly focus on controlling the emotion~\cite{EMOTE, peng2023emotalk} and style~\cite{voca, meshtalk, imitator, xing2023codetalker, fan2022faceformer} of the animation.
They cannot allow users to easily edit specific-parts of the animation sequence.
I.e., if a user wants to edit the style of a part of an existing sequence, they have to generate a new sequence with desired style and then blend it with the original. 
Such an edit is often impractical, time-consuming and requires frame-by-frame manual inspection to ensure accurate lip-sync.
Notably, diffusion-based facial motion synthesis methods ~\cite{FaceDiffuser_Stan_MIG2023, sun2023diffposetalk, aneja2023facetalk, chen2023diffusiontalker}, where such an editing could be considered as a byproduct of diffusion models, are not demonstrating this.
Further, recent works~\cite{FaceDiffuser_Stan_MIG2023, aneja2023facetalk, chen2023diffusiontalker} focus on showcasing diversity in eye-blinks and upper face motion, which have a weak (if any) correlation with the audio.
However, the ability to lip-sync animations in varied but plausible ways is also essential, especially for the animation and movie dubbing.
Because of these shortcomings of learning-based methods, procedural methods, though tedious and labor-intensive, still dominate 3D animation in movies and gaming, highlighting the need for a more efficient alternative.

\begin{figure}[t!]
    \centering
    \includegraphics[width=\columnwidth]{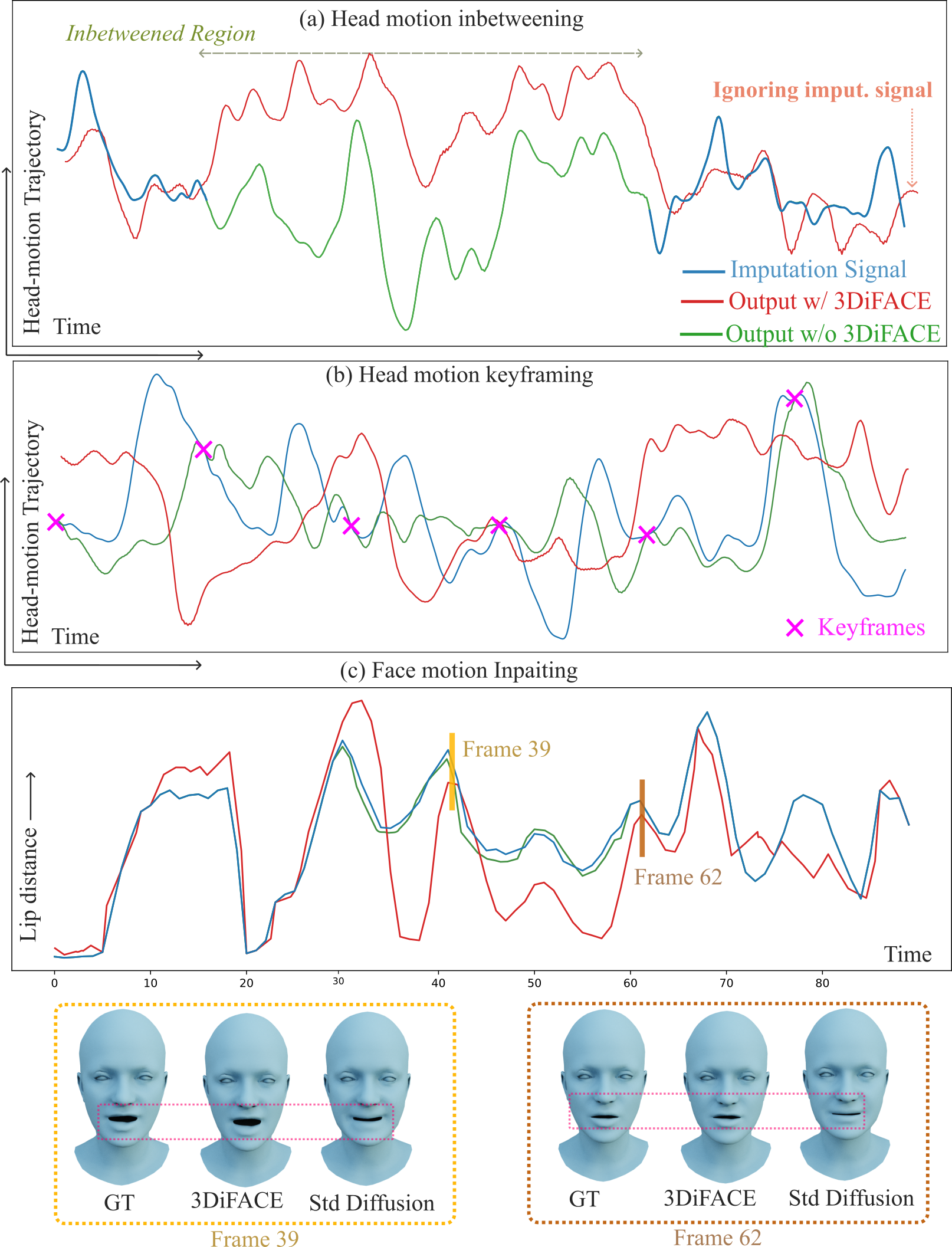} 
    \caption{Illustration of holistic 3D facial motion editing with and without \textit{3DiFACE}.
    Head motion editing is shown in (a) and (b), where one can see that standard diffusion is ignoring the imputation signal.
    Facial motion editing (c) shows the unrealistic style-shifts for classical diffusion, refer \textit{Frame 39} and \textit{Frame 53}.
    }
    \label{fig:3DiFACE_impact}
\end{figure}

Our goal is to achieve diverse motion synthesis with precise control from partial control signals like keyframes (also referred as Imputation signal). In this context, we face three main challenges: 
(i) Facial movements are highly person-specific.
If the speaking style of the edited region isn't personalized and doesn't match the imputation signal, it results in unrealistic animations due to sudden style shifts between edited and unedited motion (see~\Cref{fig:3DiFACE_impact}). 
(ii) Diffusion models are known to require large training sets \cite{rombach2022high}, yet the size of existing high-quality speech-to-3D-animation datasets is limited.
Additionally, for personalization of speaking-style the model should be capable of fine-tuning on a short reference video ($1min$).
(iii) Standard diffusion-based editing on head motion often ignores the imputation signal, as also observed in~\cite{karunratanakul2023gmd} (see~\Cref{fig:3DiFACE_impact}).

To tackle these challenges, we propose a diffusion-based method for speech-driven holistic 3D facial animation synthesis and editing.
Specifically, to address the aforementioned challenges (i) and (ii), we employ a fully convolutional 1D U-net architecture that can be trained on the small VOCAset~\cite{voca} and fine-tuned using a short $1min$ reference video of the target subject.
The fully convolutional design of our method allows to sub-divide the input sequence into $1sec$ viseme-level motion segments during training and generalize to sequences of arbitrary length at inference time. 
Specifically, we leverage the viseme-level motion diversity present in the dataset to train our method and generate diverse sequence-level samples for a single audio input at inference. 
For head motion editing (iii), we introduce a novel Sparsely-Guided Diffusion (SGDiff) approach.
This method involves replacing part of the noisy sequence with ground truth data and enforcing the model to precisely replicate the samples within the ground truth region.
This approach prevents the model from ignoring the sparse imputation signal, resulting in smoother and more natural head-motion editing.
Through quantitative, qualitative, and perceptual evaluation, we demonstrate the superiority of our method in producing diverse personalized facial animation with natural head motions.
Further, we demonstrate the importance of our architecture design choices, data-efficiency and robustness in detailed ablation studies.

\medskip\medskip\medskip
\noindent
In summary, our contributions are twofold:
\begin{itemize}
  \item  a fully convolutional speech-driven diffusion model that leverages viseme-level diversity to synthesize diverse holistic 3D facial animations of arbitrary length.
  \item employing personalization and sparsely-guided diffusion, we demonstrate explicit 3D facial animation editing, including seamless motion interpolation and keyframing.
\end{itemize}

\section{Related Work}
\label{sec:related}

\textbf{Speech-Driven 3D Facial Animation:} Methods for 3D facial animation synthesis can be classified into 2 categories namely procedural and learning-based methods.  \textit{Procedural methods:}
Earlier works on 3D facial animation used procedural rules~\cite{de_martino_facial_2006,ezzat_miketalk:_1998, kalberer_face_2001, edwards2016jali} to map audio to facial rigs.
The methods offer artists precise control and ability to modify part of the animation. Despite being manual and labor-intensive, they remain the standard for 3D animation in movies and gaming.
\textit{Learning-based methods:} using motion-captured 3D audio visual datasets like VOCA~\cite{voca} or BIWI~\cite{eth_biwi_00760}, statistical approaches~\cite{cao, gen-speech-animation, thies2020nvp, karras_audio-driven_2017, voca, meshtalk, fan2022faceformer, imitator, peng2023selftalk, EMOTE, hui2024Mimic, 10504116, 10555000} learn to animate 3D meshes or blendshapes from audio inputs.
They produce facial animation with high lip-synchronization. However they struggle to model head-motion and the many-to-many mapping between audio and facial animations.
Concurrent to our work, several methods~\cite{aneja2023facetalk, sun2023diffposetalk, FaceDiffuser_Stan_MIG2023, chen2023diffusiontalker} employ Diffusion-probabilistic models~\cite{ho2020denoising} to learn and synthesize 3D facial animation with diversity. 
DiffPoseTalk~\cite{sun2023diffposetalk} and Media2Face~\cite{zhao2024media2face} are closest to our 3D facial animation method with head-motion synthesis.
DiffPoseTalk~\cite{sun2023diffposetalk} synthesizes personalized 3D facial animation with head motion, however, it cannot offer any editing capabilities and the synthesis quality is limited by the tracker.
Media2Face~\cite{zhao2024media2face} utilizes an in-house 4D corpus over 200 subjects to learn a motion prior and subsequently builds a large in-the-wild training corpus with text annotations to train a diffusion model with various global control signals.
In contrast to the above works, our approach focuses on synthesizing personalized holistic facial animation, that can be precisely controlled using user-defined key-frames, thus, bridging the fine control from procedural methods with diverse multi-modal synthesis from learning-based methods.

Concurrent works~\cite{sung2024Multitalk, nocentini2024scantalk3dtalkingheads, sung2024laughtalk} have focused on improving the generalization to different language, removing topological constraints and synthesizing animation with authentic laughter. As this is not the focus of our work, we suggest interested  readers to check out the respective works.

\medskip
\noindent
\textbf{3D Holistic Facial motion Editing:}
We define facial motion editing as the task of explicitly editing an input sequence, by defining keyframes and regenerating selected parts of it.
3D animation artists can manually define these key-frames (e.g. from a database, or an already generated sequence) and refine them further to get desired expressions and poses at specific points in the motion sequence.

Procedural animation methods~\cite{de_martino_facial_2006,ezzat_miketalk:_1998, kalberer_face_2001, edwards2016jali} allow this precise control via modifying the animation curves in the interested parts of the sequence. However as mentioned in the previous section, they are labour intensive and limited in terms of animation styles.
Learning-based methods offer to control style~\cite{xing2023codetalker, voca, meshtalk, fan2022faceformer, imitator, peng2023selftalk, FaceDiffuser_Stan_MIG2023, aneja2023facetalk, sun2023diffposetalk} and emotion~\cite{EMOTE, chen2023diffusiontalker} during animation synthesis, however, they cannot regenerate or control part of the sequence.
Further, diffusion-based facial motion synthesis methods ~\cite{FaceDiffuser_Stan_MIG2023, sun2023diffposetalk, aneja2023facetalk, chen2023diffusiontalker}, where editing might be considered as a byproduct of diffusion models, are not demonstrating this.
Design choices such as auto-regressive mechanisms~\cite{aneja2023facetalk, FaceDiffuser_Stan_MIG2023, sun2023diffposetalk, chen2023diffusiontalker}, self-attention with look-ahead masks~\cite{aneja2023facetalk, sun2023diffposetalk}, and lack of speaking style personalization~\cite{FaceDiffuser_Stan_MIG2023, aneja2023facetalk} prevent these models from effectively editing facial motions.
Concurrent work, Media2face~\cite{zhao2024media2face} demonstrate the ability to locally edit facial animation in one of the sequence in the paper. However, several questions regarding the effectiveness, head-motion editing and adaptability to new subject remain open.
In this work, we propose a diffusion-based holistic facial animation method that can synthesize 3D holistic animation from input audio and allowing to locally edit the generated or existing animation sequence.
Further, through detailed experiments, we demonstrate the effectiveness of our method, adaptability to new subjects and ability to edit head-motions.

\medskip
\noindent
\textbf{Diffusion Guidance:}
Diffusion models have significantly advanced generating images~\cite{ho2020denoising, ho2022classifierfree, Rombach_2022_CVPR}, videos~\cite{guo2023animatediff, ho2022video, esser2023structurecontentguidedvideosynthesis, ruan2022mmdiffusion}, audio~\cite{lemercier2024diffusionmodelsaudiorestoration, evans2024fasttimingconditionedlatentaudio, wang2023auditaudioeditingfollowing}, and motion~\cite{ tevet2023human, dabral2022mofusion, xie2024omnicontrol, karunratanakul2023gmd}. 
Control in these models is implemented through several methods: Classifier-guidance~\cite{DBLP:journals/corr/abs-2105-05233} uses a classifier's gradient, while classifier-free guidance~\cite{ho2022classifierfree} balances quality and diversity with conditional and unconditional models. 
ControlNet~\cite{zhang2023adding} employs a trainable copy of a diffusion model for processing conditions, and inpainting methods~\cite{karunratanakul2023gmd} generate consistent outputs from partial data.
Guided motion diffusion(GMD)~\cite{karunratanakul2023gmd} a full-body motion synthesis method, is the work closest to ours. GMD support spatial guidance, cannot offer keyframing and control on sparse pose. 

\section{Preliminaries}
\label{sec:preliminaries}
\paragraph{Denoising Diffusion Probabilistic Models:}
Our method is based on the diffusion framework of Sohl et al.~\cite{sohl2015deep}, where a training sample \(x_0\) gradually transforms into white noise through the addition of Gaussian noise across \(T\) steps. 
Following Tevet et al.~\cite{tevet2023human}, we train a denoising model $\theta$ that can reverse the forward diffusion and estimate the original sample \(x_0\) from a noised version \(x_t\), conditioned on input $C$: $ \hat{x}_0 = \theta(x_t, t, C)$.
To generate new samples, we start from random noise \(x_T\) and apply iterative denoising until reaching \(t=0\).
To improve the diversity of the samples during inference, we employ Classifier-Free Guidance (CFG)~\cite{ho2022classifierfree} and calculate the output as a weighted sum of the conditional and unconditional prediction:
\begin{equation}
    \theta(x_t, t, C) := \theta(x_t, t, \emptyset) + s \cdot \left[\theta(x_t, t, C)-\theta(x_t, t, \emptyset)\right] ,
\end{equation}
where $s$ is the guidance scale and $\theta(x_t, t, \emptyset)$ denotes the unconditional prediction (audio conditions are set to zero).
While CFG is typically used with a guidance scale $>1$ to enhance alignment with the condition, we set it to $<1$ to increase diversity (0.5 unless specified otherwise).

\vspace{-0.4cm}
\paragraph{Audio Encoding:}
\label{sec:audio_encoding}
Similar to \cite{voca, fan2022faceformer, imitator, xing2023codetalker}, we adopt the pretrained Wav2Vec2.0~\cite{wav2vec2.0} to generate audio features from the audio signal. 
Wav2Vec2.0 uses a self-supervised learning approach to map audio to quantized feature vectors with $768$ channels. 
We resample the encoder output via linear interpolation to match the sampling rate of the motion sequences (30fps for VOCAset~\cite{voca}).
A trainable linear layer is applied to project the feature vectors to $64$ channels, resulting in a speech representation $\hat{A} \in \mathbb{R}^{N\times64}$ for $N$ frames.

\section{Method}
\label{sec:method}

\begin{figure}[h!]
    \centering
    \includegraphics[width=\columnwidth]{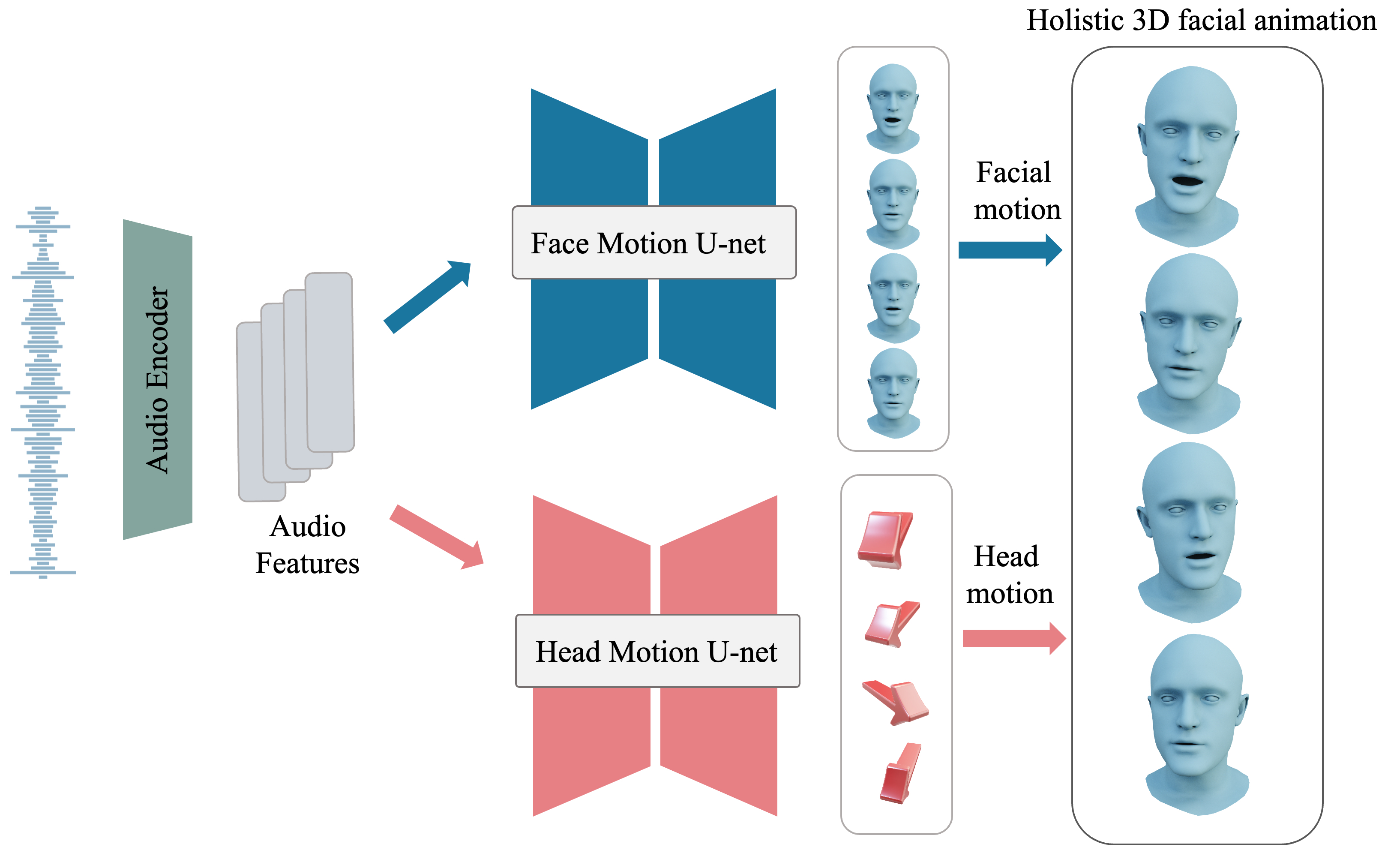} 
    \caption{Overview of our method.
    We employ two diffusion-based motion generators with shared audio encoder to model 3D facial and head motion separately.
     }
     \vspace{-0.1cm}
    \label{fig:overview}
\end{figure}

Our goal is to synthesize and edit holistic 3D facial animation given an input audio signal. 
In this context, `holistic' refers to animation with facial \textit{and} head motion, which we model in two diffusion-based networks, see \Cref{fig:overview}.
This is motivated by the fact that the facial motion is highly correlated to the speech signal, while the correlation w.r.t. the head motion is weaker and, thus, requires a longer context of information, hence, a different training scheme (and data).
The two diffusion models for facial ($\theta_f$) and head ($\theta_h$) motion are conditioned on the encoded audio signal $\hat{A}$ using a pretrained Wav2Vec2.0~\cite{wav2vec2.0} as explained in \Cref{sec:preliminaries}.
We leverage convolutional architectures for the denoising models which we describe in the following.

\subsection{Facial motion generator}
\label{face_generator}

\begin{figure}[t]
    \centering
    \vspace{-1.2cm}
    \setlength{\unitlength}{0.1\linewidth}
    \begin{picture}(10, 6)
    \put(0, 0){\includegraphics[width=1.03\linewidth]{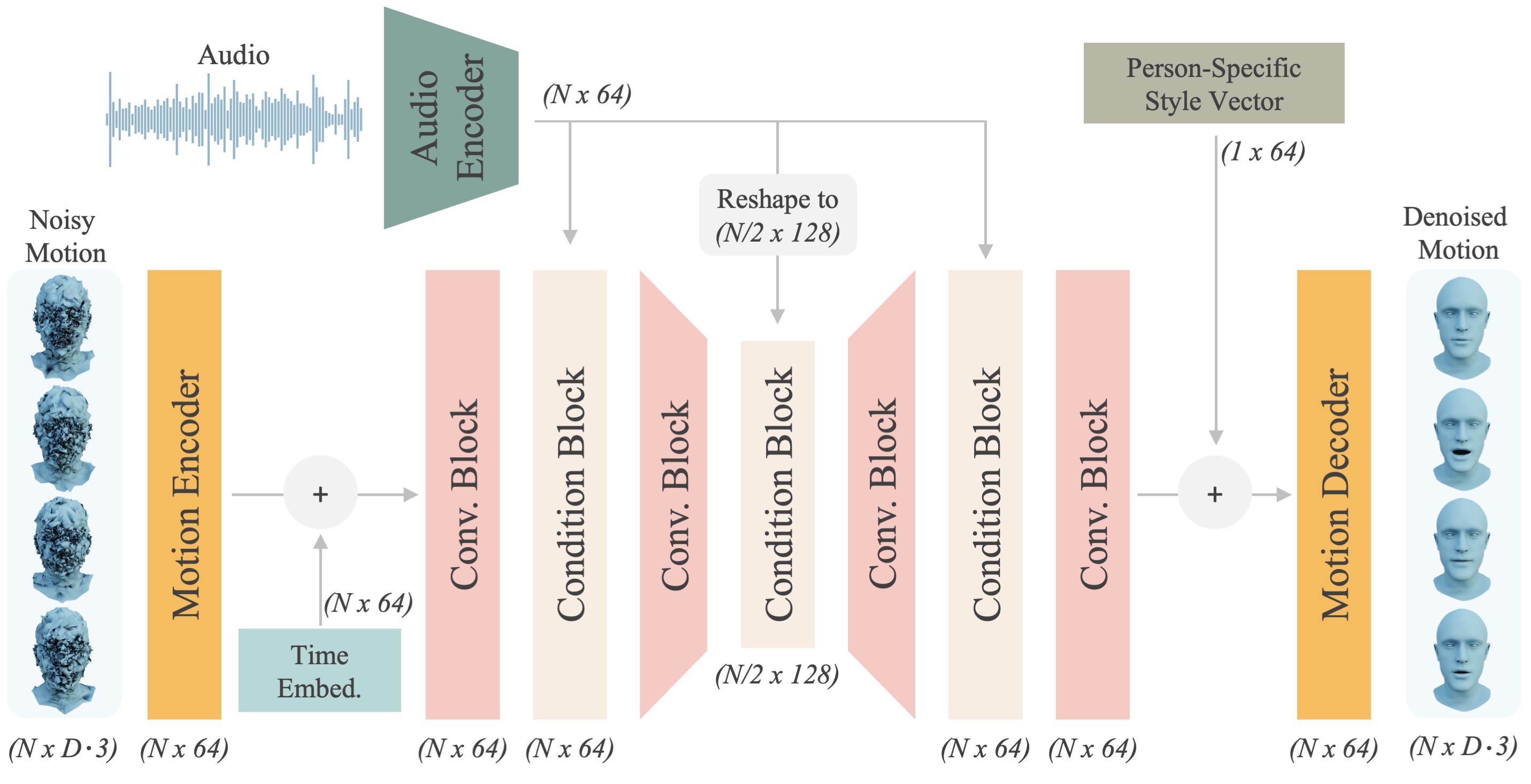}}
    \put(3.85, 4.8){\fontsize{2}{4}\selectfont $\hat{A}$}
    \put(0.3, 4.0){\fontsize{2}{4}\selectfont $x_t$}
    \put(9.7, 4.0){\fontsize{2}{4}\selectfont $\hat{x}_0$}
    \put(8.25, 3.92){\fontsize{1.0}{2}\selectfont $S_i$}
    \end{picture}
    \vspace{-0.7cm}
  \caption{Our facial motion generator takes noised vertex displacements, denoted as $x_t$, and the diffusion time step embedding as inputs to predict a denoised sample $\hat{x}_0$, leveraging both the audio features signal $\hat{A}$ and a person-specific feature vector $S_i$.
Note that $N$ corresponds to the frame count of the sequence and $D$ to the number of vertices. 
}
\label{fig:method}
    \vspace{-0.3cm}
\end{figure}

Our diffusion-based facial motion generator takes an audio signal as input and produces a sequence of 3D vertex displacements w.r.t. a template mesh by iterative denoising, see \Cref{fig:method}.
Let $x_0\in \mathbb{R}^{N\times D \cdot 3}$ denote such a sequence of displacements, where $N$ is the sequence length and $D$ is the number of vertices in the template mesh. 
The input to our diffusion model parameterized by $\theta_f$ is a noisy vertex displacement sequence $x_t\in\mathbb{R}^{N\times D \cdot 3}$. The task is then to predict its noise-free counterpart $\hat{x}_0=\theta_f(x_t, t, C_f)$, given diffusion step $t$ and conditions $C_f$. 
Note that in our formulation, the condition $C_f$ represents the set of both the per-frame audio features $\hat{A}$ and person-specific feature $S_i$.

In contrast to state-of-the-art methods on 3D facial animation synthesis that utilize transformer architectures~\cite{fan2022faceformer, imitator, xing2023codetalker, sun2023diffposetalk},
we adopt a 1D-convolution network inspired from Pavllo et al.~\cite{pavllo20193d}.
Specifically, we replace the commonly used attention-based condition injection with feature concatenation. 
Our fully convolutional architecture, free from attention, allows us to sub-divide the input sequence into viseme-level motion segments (e.g., $30$ frames) during training and to generalize to sequences of arbitrary length at inference time. 
We empirically observed that these modifications to the architecture are critical to train on the limited VOCA training dataset \cite{voca}, especially in the unconditional training setup (see \Cref{tab:ablation}, row 1-4). %
Note that this strategy is not viable for transformer-based 3D facial animation baselines, since it struggles to capture any longer-term dependency beyond the predefined context length, leading to context fragmentation~\cite{Dai2019TransformerXLAL} and subpar performance (see \Cref{tab:ablation}, row 3).
This issue becomes even more pronounced in our training setting, where the sequences have only $30$ frames.
While auto-regressive motion synthesis could in theory mitigate this limitation, it would make the animation editing tasks, such as motion inbetweeing, impossible. 

\subsubsection{Training} 
\label{head_motion_training}
We train our model to predict the vertex displacements $x_0$ from their noised counterparts $x_t$ on VOCAset~\cite{voca}: 
\begin{equation}
 \mathcal{L}_{\text{simple}} = || x_{0} - \theta_f(x_{t},t,C_f) || ^2 .
\end{equation}
In contrast to predicting the applied noise which is common practice \cite{ma2022mofusion, zhang2022motiondiffuse, rombach2022high}, we empirically found that predicting the ground truth displacements yields better convergence in the unconditional and person-specific fine-tuning case. 
To improve temporal smoothness \cite{imitator}, we add a velocity loss:
\begin{equation}
 \mathcal{L}_{\text{vel}} = \dfrac{1}{N-1} \sum_{n=1}^{N} || (x_{0,n} - x_{0, n-1}) - (\hat{x}_{0,n} - \hat{x}_{0,n-1}) || ^2 ,
\end{equation}
where $x_{0,n}$ denotes the ground truth vertex displacements in frame $n$.
Our final training objective is formulated as:
\begin{equation}
    \label{eq:face}
 \mathcal{L}_{\text{face}} = \mathcal{L}_{\text{simple}}  + \lambda_{\text{vel}}\cdot\mathcal{L}_{\text{vel}} ,
\end{equation}
with $\lambda_{\text{vel}}=10.0$. %
Note that during training, we randomly set the audio features $\hat{A}$ in the condition $C_f$ to 0 for 10\% of the time to enable unconditional synthesis at inference time.

\subsubsection{Person-specific fine-tuning}
\label{personalized_fine_tuning}
As described in the introduction, personalization of speaking-style is indispensable for facial motion editing.
For capturing the speaking style of a subject that is not part of the training set, we require a short reference video. The facial movements are reconstructed with the state-of-the-art monocular face tracker MICA~\cite{mica}. 
We use the reconstructed meshes as pseudo ground truth and fine-tune the entire model to fit the expression distribution of the target subject using the training objective from Eq.~\eqref{eq:face}.

\subsection{Head-motion generator with sparse guidance}
\label{head_generator}
\begin{figure}[t]
    \centering
    \includegraphics[width=\linewidth]{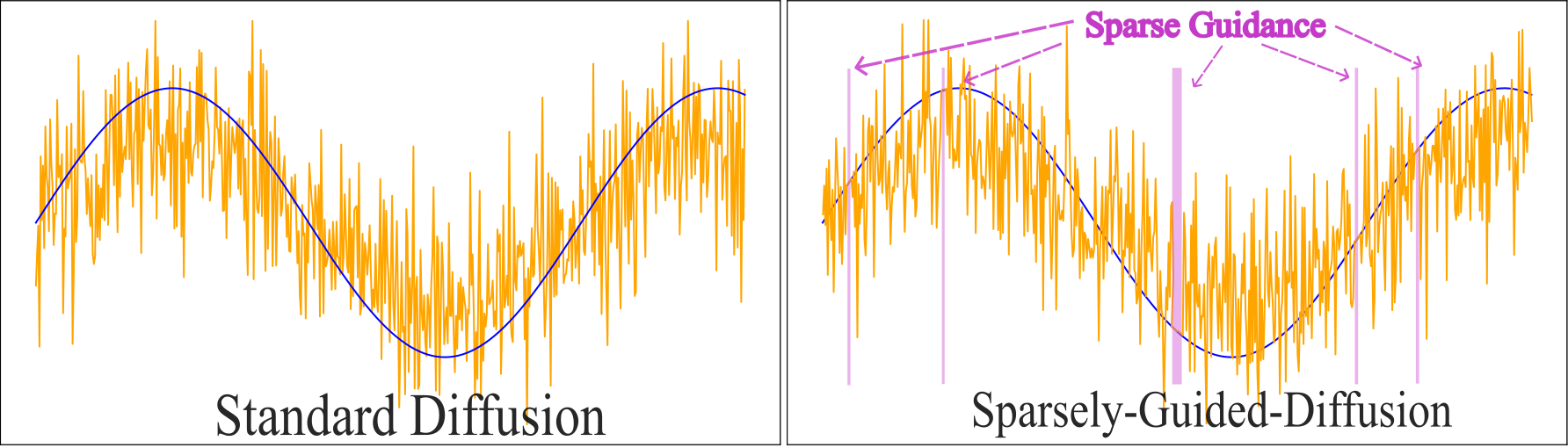} 
    \caption{Illustration of standard diffusion (left) and our sparsely-guided diffusion (right), where in the forward diffusion process, part of the noisy input signal is replaced with the ground truth signal and a guidance flag of (0) and (1) is concatenated to the noisy and ground truth regions respectively.
    }
    \label{fig:std_vs_gmd_comp}
    \vspace{-0.3cm}
\end{figure}

Given an audio signal input, our head motion generator produces smooth and natural head motions $y_0\in \mathbb{R}^{N\times 3}$, where $N$ is the sequence length.
We parameterize the head motion via the neck joint rotation in the FLAME model~\cite{flame}, where the rotation is represented via axis angle. %
Motivated by the head-motion editing issue mentioned in the introduction, we introduce a sparsely-guided diffusion (SGDiff) for the head motion synthesis. 
Specifically, in addition to the audio features, we inject an intra-sequence guidance to highlight the relative importance of the different segments in the input signal.
As illustrated in \Cref{fig:std_vs_gmd_comp}, during the forward diffusion process, part of the noisy input ($y_t$) is replaced with ground truth signals, and a corresponding guidance flag of 0 or 1 (ground truth signal) is concatenated.
A denoising model parameterized by $\theta_h$ is trained to reverse this diffusion process by leveraging this additional information.

Similar to the facial motion generator, we employ a fully convolutional architecture as our backbone for the head-motion denoising model.
Additionally, we introduce skip connections between the encoder and decoder layers, to aid the model in reproducing the sparse ground truth signals.
For the audio encoding, we use the pre-trained audio encoder from the facial motion synthesis pipeline, which is kept frozen during the head-motion training.
The final diffusion formulation for diffusion step $t$ is $\hat{y}_0=\theta_h(y_t, t, \hat{A})$.

\subsubsection{Training}
The complete training procedure of the sparsely-guided diffusion is detailed in~\Cref{algo:modified_gmd_train}. %
In addition to the losses used in the facial motion generator (\Cref{head_motion_training}), we add an additional guiding mask loss to enforce the model to faithfully reproduce the results of the ground truth signal injected into the sequence. 
The guidance loss is:
\begin{equation}
 \mathcal{L}_{\text{mask}} =  || w_{0,n} \odot (y_{0,n} - \theta_h (y_{t,n},t, \hat{A})) || ^2 ,
\end{equation}
where $n$ indicates the $n^{th}$ frame in sequence $y_0$, $w_{0,n}$ is the guidance weight, $1$ for ground truth frames and zero otherwise and $\odot$ is the Hadamard product.

\begin{algorithm}
\caption{Our SGDiff Training}
\label{algo:modified_gmd_train}
\footnotesize 
\begin{algorithmic}[1]
\Repeat
\State $y_0 \sim q(y_0)$ \hfill \# sample from train distribution
\State $t \sim \text{Uniform}(\{1, \ldots, T\})$
\State $\epsilon \sim \mathcal{N}(0, I)$
\State $y_t = \sqrt{\bar{\alpha}_t} y_0 + \sqrt{1-\bar{\alpha}_t} \epsilon$  \hfill \# $\bar{\alpha}_t$ denotes diffusion noise schedule
\State $\bar{y_t} = y_t \oplus (0) $ \hfill \# $\oplus$ = concatenation operation
\State $\bar{y_0} = y_0 \oplus (1) $  
\State $y_t = (1-M_t) \odot \bar{y_t} + M_t \odot \bar{y_0} $ 
  \hfill  \# Guidance injection
\State grad desc. $\nabla_{\theta_h} \left\| y_0  - \theta_h (y_t, t, \hat{A}) \right\|^2$
\Until{converged}
\State \# $M_t$ = random imputation mask
\end{algorithmic}
\end{algorithm}

\begin{algorithm}
\caption{Our SGDiff Sampling}
\label{algo:modified_gmd_sampling}
\footnotesize 
\begin{algorithmic}[1]
\State Input signal $Y_0$, if any
\State Imputation mask $M_0$, if any
\State $y_T \sim \mathcal{N}(0, I)$
\State $\bar{Y_{0}} = Y_0 \oplus (1)$
\For{$t = T, \ldots, 1$}
\State $\bar{y_{t}} = y_{t} \oplus (0)$
\State $y_{t} = (1-M_0) \odot \bar{y_{t}} + M_0 \odot \bar{Y
_0} $
\State $\hat{y_0} = \theta_h(y_t, t, \hat{A})$
\State $\hat{y_0} = (1-M_0) \odot \hat{y_0} + M_0 \odot Y_0 $
\State $\mu, \sigma \gets \mu(y_{t}, \hat{y_0}), \sigma_t$ 
\State $y_{t-1} \sim \mathcal{N}(\mu, \sigma)$
\EndFor
\State \Return $y_0$
\end{algorithmic}
\end{algorithm}

\subsection{Sampling and editing of a 3D facial animation}
Following standard diffusion methods~\cite{ho2020denoising, tevet2023human}, we generate new samples by starting from random noise ($X_T$ for facial motion and $Y_T$ for head motion) drawn from Gaussian noise and iteratively denoise for $T$ steps using the respective denoising models $\theta_f$ and $\theta_h$. 
For control and part-regeneration, we replace the corresponding noisy sample ($x_t$ or $y_t$) with the imputation signal at each time $t$ before denoising.
For facial motion editing, we personalize $\theta_f$ to the target subject using the steps detailed in~\Cref{personalized_fine_tuning} before iterative denoising. 
For head motion editing, we follow the procedure in~\Cref{algo:modified_gmd_sampling}, replacing the input ($y_t$) with the imputation signal and adding a guidance flag.

\section{Dataset}
\label{sec:data}
We train our facial motion model on VOCAset~\cite{voca}, since it provides high-quality, speech-aligned 3D face scan sequences. 
Following previous works~\cite{fan2022faceformer, imitator, xing2023codetalker}, we use the train/val/test set split of $8,2,2$ actors.
All $40$ sequences of the training actors are used during training. 
However, for the test and validation, only $20$ sequences without overlap with the speech scripts of the training sequences are used.
We evaluate person-specific fine-tuning on in-the-wild videos from Imitator~\cite{imitator}. The provided videos are 2 minutes long which we divide into 60/30/30 seconds for train/val/test respectively.
To train our head-motion generator, we use the HDTF~\cite{HDTF} dataset, as the VOCAset does not include head motion.
Using the download and processing script provided by the authors, we extract $352$ videos with 246 unique subjects and use the MICA tracker~\cite{mica} to extract head poses.
For our experiments, we split the dataset into 300/20/32 sequences for train/val/test accordingly. 
We employ the VOCAset, HDTF, and Imitator's in-the-wild dataset to train our method for generating and editing 3D facial animations with head-motion. 
This choice led us to exclude the Biwi dataset~\cite{eth_biwi_00760} from our study, as it lacks sequences with full head model like FLAME~\cite{flame}, which is essential for synthesizing head motion effectively.
For more details refer the supplemental document.

\begin{figure*}[t]
    \centering
       \resizebox{0.65\textwidth}{!}{
        \begingroup%
  \makeatletter%
  \providecommand\color[2][]{%
    \errmessage{(Inkscape) Color is used for the text in Inkscape, but the package 'color.sty' is not loaded}%
    \renewcommand\color[2][]{}%
  }%
  \providecommand\transparent[1]{%
    \errmessage{(Inkscape) Transparency is used (non-zero) for the text in Inkscape, but the package 'transparent.sty' is not loaded}%
    \renewcommand\transparent[1]{}%
  }%
  \providecommand\rotatebox[2]{#2}%
  \newcommand*\fsize{\dimexpr\f@size pt\relax}%
  \newcommand*\lineheight[1]{\fontsize{\fsize}{#1\fsize}\selectfont}%
  \ifx\svgwidth\undefined%
    \setlength{\unitlength}{600.88976378bp}%
    \ifx\svgscale\undefined%
      \relax%
    \else%
      \setlength{\unitlength}{\unitlength * \real{\svgscale}}%
    \fi%
  \else%
    \setlength{\unitlength}{\svgwidth}%
  \fi%
  \global\let\svgwidth\undefined%
  \global\let\svgscale\undefined%
  \makeatother%
  \begin{picture}(1,0.70707071)%
    \lineheight{1}%
    \setlength\tabcolsep{0pt}%
    \put(-0.01,0.55010475){\rotatebox{90}{\makebox(0,0)[lt]{\lineheight{1.25}\smash{\begin{tabular}[t]{l}Ours: sample 1\end{tabular}}}}}%
    \put(-0.01,0.39444623){\rotatebox{90}{\makebox(0,0)[lt]{\lineheight{1.25}\smash{\begin{tabular}[t]{l}Ours: sample 2\end{tabular}}}}}%
    \put(-0.01,0.2059717){\rotatebox{90}{\makebox(0,0)[lt]{\lineheight{1.25}\smash{\begin{tabular}[t]{l}TalkSHOW~\cite{yi2023generating}\end{tabular}}}}}%
    \put(-0.01,0.04264139){\rotatebox{90}{\makebox(0,0)[lt]{\lineheight{1.25}\smash{\begin{tabular}[t]{l}SadTalker~\cite{sadtalker}\end{tabular}}}}}%
    \put(0,0){\includegraphics[width=\unitlength,page=1]{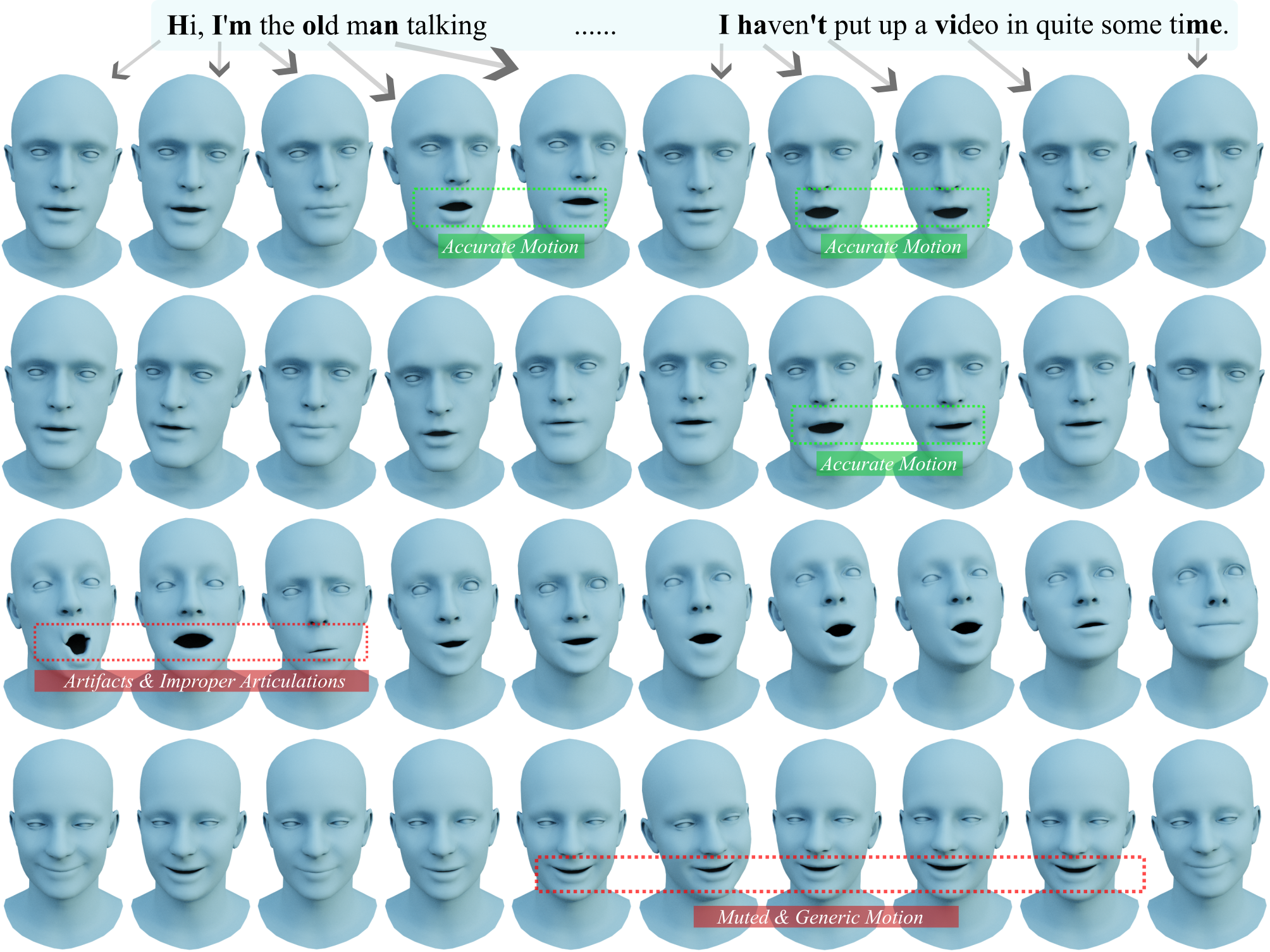}}%
  \end{picture}%
\endgroup%

        }
    \caption{
    Qualitative comparison: Our method outperforms the baseline in creating more accurate lip-synced facial animations with diverse head movements. Specifically, TalkSHOW produces animations with jittery artifacts, while SadTalker yields muted and generic animations.
    }
    \label{fig:holistic_qual_comp}
\end{figure*}

\section{Results}
\label{sec:results}

We evaluate our method against state-of-the-art methods: SadTalker~\cite{sadtalker} and TalkShow~\cite{yi2023generating} on the holistic 3D facial animation synthesis task and VOCA~\cite{voca}, Faceformer~\cite{fan2022faceformer}, CodeTalker~\cite{xing2023codetalker}, EMOTE~\cite{EMOTE}, FaceDiffuser~\cite{FaceDiffuser_Stan_MIG2023} and Imitator~\cite{imitator} on facial motion synthesis task.
\Cref{fig:holistic_qual_comp} presents the qualitative comparison on holistic 3D motion synthesis, where our method produces more accurate lip-synced facial animations with diverse head movements.
Additional qualitative results are shown in the suppl. material and video.

\medskip
\noindent
\textbf{Quantitative Comparison:} 
\begin{table}[t!]
    \centering
\resizebox{\linewidth}{!}{
    \begin{tabular}{cl|cccc} \toprule
         &\textbf{Method} & $\mathbf{Div^{L}}$ $\uparrow$  & \textbf{Lip-Sync} $\downarrow$  & $\mathbf{BA}$ $\uparrow$ & $\mathbf{Div^{H}}$ $\uparrow$ \\
         \midrule
            & & \multicolumn{4}{c}{Holistic 3D Facial animation syn.} \\ 
            \cmidrule(r){3-6} 
            1 & SadTalker~\cite{sadtalker} & $1.59$ & $4.01$ & $0.285$ &  $0.004$\\
            2 & TalkSHOW~\cite{yi2023generating} & $1.80$ & $4.35$ & $0.296$ &  $0.002$\\
            3 & \textbf{Ours composite}  & $\mathbf{2.57}$ & $\mathbf{1.71}$  & $\mathbf{0.338}$ &  $\mathbf{0.007}$\\
            \midrule
             & &  \multicolumn{4}{c}{Non-Personalized regression} \\ 
            \cmidrule(r){3-6} 
            4 & VOCA~\cite{voca} & $-$ & $5.30$ &  $-$ &  $-$ \\
            5 & Faceformer~\cite{fan2022faceformer} & $-$ & $2.85$ &   $-$ &  $-$ \\
            6 & Imitator~\cite{imitator} & $-$ & $1.95$  & $-$ &  $-$ \\
            7 & CodeTalker~\cite{xing2023codetalker} & $1.40$  & $2.55$ & $-$ &  $-$ \\
            8 & Ours$_{s=0.5}$ (w/o sty) & $\mathbf{2.57}$ & $\mathbf{1.71}$ & $-$ &  $-$ \\
            \midrule
             & & \multicolumn{4}{c}{Non-Personalized diffusion} \\ 
             \cmidrule(r){3-6} 
            9 & FaceDiffuser~\cite{FaceDiffuser_Stan_MIG2023} & $0.05$ &  $\mathbf{1.60}$ & $-$ &  $-$ \\
            10 & Ours$_{s=1.0}$ (w/o sty) & $\mathbf{0.64}$ & $1.62$  & $-$ &  $-$ \\
            \midrule
             & & \multicolumn{4}{c}{Personalized synthesis} \\ 
            \cmidrule(r){3-6} 
            11 & Imitator (w/ sty) & $-$ & $\mathbf{1.35}$  & $-$ &  $-$ \\
            12 & Ours$_{s=0.5}$  (w/ sty) & $\mathbf{1.57}$ & $1.56$  & $-$ &  $-$ \\
            13 & Ours$_{s=1.0}$  (w/ sty) & $0.24$ & $1.42$  & $-$ &  $-$ \\
            \midrule
    \end{tabular}
}
\caption{ Quantitative comparison: 
Our proposed method produces better holistic 3D facial animations with high-fidelity lip and head motions (refer row 1-3). 
On the non-personalized regression and diffusion facial motion synthesis task (row 4-10), our method produces outperforms the baselines, except for FaceDiffuser, where we match the performance on \textit{Lip-Sync} despite producing more diverse samples.
Finally, our method is able to personalize facial motions on the level of Imitator~\cite{imitator}, while producing more diverse samples and allowing for motion editing using keyframes.
}
\label{tab:exprs_quan_study}
\vspace{-0.2cm}

\end{table}
In \Cref{tab:exprs_quan_study}, we present a quantitative evaluation based on the following metrics: 
\textit{Lip-Sync} measures the lip synchronization using Dynamic Time Warping to compute the temporal similarity~\cite{imitator}.
Diversity metric $Div^L$ and $Div^H$ proposed by Ren et al.~\cite{ren2023diffusion} measures the diversity of lip motion and head motion generated from the same audio.
Similar to DiffPoseTalk~\cite{sun2023diffposetalk}, we employ a modified beat alignment \textit{BA} to measure the synchronization of the head movement beats. %
Please refer the suppl. material for detailed information about the metrics.

From ~\Cref{tab:exprs_quan_study} (rows 1-3), we see that our method 
significantly surpasses the baselines in \textit{holistic 3D facial animation synthesis}, particularly in terms of lip-sync accuracy and beat alignment, while offering greater diversity.

For the \textit{facial motion synthesis task without head motion}, we quantitatively compare our method on three different setups namely, non-personalized regression and diffusion and personalized synthesis.
In the non-personalized regression and diffusion setup, our method outperforms the baselines, except for FaceDiffuser, where we match the performance on \textit{Lip-Sync} despite producing more diverse samples (refer to~\Cref{tab:exprs_quan_study} rows 4-10).
Note that we can adjust the guidance scale parameter to control synthesis diversity and lip-sync accuracy, which FaceDiffuser~\cite{FaceDiffuser_Stan_MIG2023} cannot do. Please refer the supplemental document for a more detailed study on the impact of guidance scale $s$. 
Finally, in the \textit{personalization synthesis} setup, we achieve higher synthesis diversity compared to Imitator~\cite{imitator} and match the performance closely in terms of \textit{Lip-Sync} (refer to~\Cref{tab:exprs_quan_study} rows 11-13).
Note that Imitator is a deterministic model that does not allow for diverse lip-motion synthesis and facial motion editing.

\medskip
\noindent
\textbf{User Study:}
\begin{table}
     \resizebox{\columnwidth}{!}{
    \begin{tabular}{cl|cc} \toprule
        & &  \multicolumn{2}{c}{Holistic synthesis} \\ 
        \cmidrule(r){3-4} 
       &\textbf{Method} & \textbf{Face Motion} (\%) &  \textbf{Head motion} (\%) \\ \midrule
        1  & Ours vs SadTalker~\cite{sadtalker}                         & $88.13$ & $86.43$ \\
        2  & Ours vs TalkShow~\cite{yi2023generating}                   & $90.77$ & $87.96$ \\
         \midrule
        &\textbf{Method} & \textbf{Exprs} (\%) &  \textbf{Lip-sync} (\%) \\ \midrule
        & &  \multicolumn{2}{c}{High-Fidelity (Ours $s=1.0$)} \\ 
        \cmidrule(r){3-4} 
        3 & Ours  vs Imitator~\cite{imitator}                      & $65.72$ & $69.47$ \\
        4 & Ours vs Faceformer~\cite{fan2022faceformer}                  & $73.28$ & $71.43$ \\
        5 & Ours  vs FaceDiffuser~\cite{FaceDiffuser_Stan_MIG2023} & $67.85$ & $66.71$ \\
        \midrule
        & &  \multicolumn{2}{c}{High-diversity (Ours $s=0.5$)} \\ 
        \cmidrule(r){3-4} 
        6 & Ours  vs CodeTalker~\cite{xing2023codetalker}                  & $53.64$ & $53.80$ \\
        7 & Ours  vs FaceDiffuser~\cite{FaceDiffuser_Stan_MIG2023} & $40.84$ & $41.55$ \\

        \bottomrule
    \end{tabular}
    }
    \captionof{table}{
    \textit{User study on holistic motion synthesis and facial motion synthesis task:}
    Compared to the baselines, our method produces consistently better holistic 3D facial animation in both low diversity($s=0.5$) and high-fidelity setup($s=1.0$).
     Similar to Imitator~\cite{imitator}, we evaluate the person-specific speaking-style similarity against~\cite{imitator}, where $55\%$ of users favored our method. %
    }
    \label{tab:perceptual_study}
\end{table}
We conducted A/B user studies to assess our method's perceptual performance.
From \Cref{tab:perceptual_study} (row 1-2), we see that our method outperforms the baselines on the holistic 3D facial animation synthesis. %
For the facial motion synthesis task, we compare our method in a \textit{high diversity} ($s=0.5$) and \textit{high fidelity} ($s=1.0$) setup.
In the high fidelity setup, we outperform the baselines in terms of both expressiveness and lip-synchronization.
Even in the high diversity setup, we outperform CodeTalker~\cite{xing2023codetalker} and perform closely to FaceDiffuser~\cite{FaceDiffuser_Stan_MIG2023}, which trades fidelity for diversity.
Furthermore, we assessed style-personalization similar to Imitator~\cite{imitator}. 
The users rated the style-similarity based on a reference video and the synthesized videos on the VOCA test set, where 55\% of the users preferred our method.
For more details on the user-study, see suppl. mat.

\medskip
\noindent        
\textbf{Motion Editing:}
To demonstrate head-motion editing, we perform both keyframing and inbetweening on head-motion data and present the results in the~\Cref{fig:3DiFACE_impact} (a) and (b).
From which we infer that our sparsely-guided motion diffusion matches the imputation signal in both, the keyframing and inbetweening scenario and produces realistic motion in the edited/re-generated part of the sequence.
Similarly, for facial motion editing, our method is able to match the speaking-style of the target imputation signal and produce smoothly edited sequences (refer ~\Cref{fig:3DiFACE_impact} (c)).
We show additional examples and a unconditional motion synthesis and editing results in the suppl. mat. and video.

\begin{table}[t!]
    \centering
\resizebox{\linewidth}{!}{%
    \begin{tabular}{cl|cc} \toprule
         &\textbf{Method} & $\mathbf{Div^{L}}$ $\uparrow$ & \textbf{Lip-Sync} $\downarrow$ \\ \midrule
           & &  \multicolumn{2}{c}{(a) Design choices} \\ 
            \cmidrule(r){3-4} 
            1 & Ours (concat + win30) & $\mathbf{2.57}$ & $\mathbf{1.71}$  \\ 
            2 & Ours (attn + win30) & $0$ & $3.21$   \\
            3 & Ours (FF arch + win30) & 0 &  $3.49$ \\
            4 & Ours (concat + no win) & $0$  & $1.98$  \\
            \midrule
            & &  \multicolumn{2}{c}{(b) Person-specific Fine-tuning} \\ 
            \cmidrule(r){3-4} 
            5 & Ours ($\sim$ 5s) & $29.95$ & $4.89$  \\
            6 & Ours ($\sim$ 30s) & $0.18$  & $1.81$ \\
            7 & Ours ($\sim$ 60s) & $0.67$  & $1.69$ \\
            8 & Ours ($\sim$100s) & $1.57$  & $1.56$ \\
           
            \midrule
             & &  \multicolumn{2}{c}{(c) GMD ablation} \\ 
            \midrule
             &\textbf{Method} & $\mathbf{BA}$ $\uparrow$  & $\mathbf{Div^{H}}$ $\uparrow$  \\ \midrule
            9 & Ours w. In mask  & $0.368$ &  $0.008$\\
            10 & Ours w. KF mask & $0.308$ &  $0.008$\\
            11 &  Ours w/o. mask & $0.338$ &  $0.007$\\
        \bottomrule
    \end{tabular}
}
\caption{
Ablation study:
(a) \textit{Design choices} study on the VOCAset~\cite{voca}  shows the importance of a fully convolutional architecture and viseme-level training.
(b) \textit{Fine-tuning Data requirement:} $30$s of video suffice to perform person-specific fine-tuning while $100$s further improve all scores (row 5-9). 
(c) \textit{GMD ablation:} shows the performance w.r.t the keyframing(sparse) and inbetweeen(dense) based  imputation signal.
}
\label{tab:ablation}

\end{table}
\medskip
\noindent
\textbf{Ablation:} In the following, we will address important questions regarding our design choices and robustness.

\medskip
\noindent
$\bullet$ 
\textit{Is a 1D-convolutional U-net architecture the right choice?} As discussed in ~\Cref{sec:method}, using our proposed architecture instead of the transformer-based architecture from Faceformer (FF) or the attention-based Unet architectures used in~\cite{ma2022mofusion} results in significantly better performance on both the \textit{Lip-Sync} and diversity (refer to~\Cref{tab:ablation} rows 1-3).

\medskip
\noindent
$\bullet$ 
\textit{What is the effect of viseme-level window-based training?} ~\Cref{tab:ablation} row 1 vs 4 shows that without window-based training the performance worsens in terms of both lip-sync and diversity.
Further, in the suppl. video, we demonstrate the ability to generate 20 $sec$ long motion compared to the baselines, despite being trained only on 1 $sec$ segments.

\medskip
\noindent
$\bullet$ 
\textit{How much data do we need for person-specific fine-tuning?}
~\Cref{tab:ablation} rows 5-8 indicate $30$ and $60$ seconds of data are sufficient for good results, $100$ seconds yield the best lip-sync and diversity $Div^L$.

\medskip
\noindent
$\bullet$ \textit{Does the sparsely-guided motion-diffusion help to generate diverse motion?}
In \Cref{tab:ablation} rows 12-14, we analyze the effect of our keyframe (KF mask) and inbetweening (In mask) based guidance on the synthesis quality. It demonstrates that employing sparsely-guided motion-diffusion improves the diversity of head-motion with minimal impact on overall quality, while offering additional editing capabilities.

\medskip
\noindent
$\bullet$  \textit{Is the performance of motion editing consistent across varying degrees of imputation signal?}
\begin{table}[t!]
    \centering
    \resizebox{\columnwidth}{!}{%
        \begin{tabular}{cl|ccccc} \toprule
             &\textbf{Method} & $\mathbf{Div^{L}}$ $\uparrow$  & \textbf{Lip-Sync} $\downarrow$ & \textbf{BA} $\uparrow$ & $\mathbf{Div^{H}}$  $\uparrow$ \\ \midrule
                1 & Ours (synthesis) & $1.35$ & $1.4$ &  $0.338$ & $0.007$  \\
                \midrule
                2 & Ours (Ip 5\%)  & $1.27$ & $1.17$ &   $0.341$ & $0.007$   \\
                3 & Ours (Ip 10\%) & $1.24$  & $1.15$ &  $0.352$  &$0.006$  \\
                4 & Ours (Ip 20\%) & $1.15$  & $1.01$ &  $0.358$  & $0.005$  \\
                5 & Ours (Ip 50\%) & $0.9$   & $0.68$ &  $0.403$ & $0.004$ \\
                \midrule
                6 & Ours (1KF/sec) & $1.26$ & $1.28$ &   $0.321$  & $0.006$ \\
                7  & Ours (2KF/sec) & $1.14$ & $1.2$ &  $0.347$ & $0.006$ \\
                8 & Ours (3KF/sec) & $1.05$ & $1.1$ &  $0.365$ & $0.005$  \\
            \bottomrule
        \end{tabular}
    }%
    \caption{
    Evaluation of editing on subject $024$ in VOCAset~\cite{voca} and HDTF~\cite{HDTF} by varying the imputation signal. %
    From the table, we observe significant improvements in synthesis quality with increase in imputation signal, indicating that the model closely matches the imputation signal and produces realistic motion.
    }
    \label{tab:motion_quan_study}
\end{table}
We evaluate the robustness of motion editing with respect to the imputation signal in both the inbetweening and keyframing scenario.
To this end, we preserve  5\%, 10\%, 20\%, and 50\% of the starting and ending frames, and then perform inbetweening for the intermediate motion sequences. 
Further, we randomly insert keyframes at different rates: 1KF/sec, 2KF/sec, and 3KF/sec and fill the motion with our method.
For facial motion, these evaluations are conducted for all sequences of the test subject 024 from the VOCAset~\cite{voca}, and the resulting metrics are presented in \Cref{tab:motion_quan_study}.
Similarly for head motion, these evaluations are conducted in the HDTF~\cite{HDTF} test set.
For the \Cref{tab:motion_quan_study}, it is evident that as the imputation signal strength increases, the synthesis's fidelity improves, indicating that model matches the imputation signal and produces realistic motion. 
This allows animators to insert any number of keyframes for fine-grained control.

\section{Discussion}
\label{sec:discussion}
Our proposed method excels in synthesizing and editing diverse holistic 3D facial animations based on speech.
Similar to \cite{imitator}, for personalization, our method depends on the quality of the face tracker.
However, through qualitative results, we demonstrate that our method is able to personalize from both high-quality motion capture sequence from VOCAset and monocular head trackers applied to in-the-wild videos.
In this work, we employ a sparsely-guided motion diffusion to tackle the imputation signal neglect in the head-motion synthesis and editing.
In contrast to head motion, for face motion editing, style-personalization is the critical contribution to enable seamless editing. 
For completeness, we include an experiment evaluating the personalization of head-motion and sparse-guidance for facial motion training in the supplemental documenet.
One key capability of our method is that it offers control to animators and creators via keyframes, which can be additionally extended to an explicit natural language based condition to control the synthesis, which we leave for future works.

\section{Conclusion}
\label{sec:conclusion}
With 3DiFACE, we presented a method that can both generate and edit diverse holistic 3D facial animations from speech input.
Through detailed experiments, we demonstrated precise control and ability to edit parts of an animation sequence.
Our work combines the precise control from procedural methods with the diverse multi-modal synthesis from learning-based methods.
We believe that these properties will make 3DiFACE a powerful tool that can reduce production time and costs, making high-quality animations more accessible, helping create lifelike avatars for movies, games, and VR, expanding creative possibilities.

\section{Acknowledgements}
This project has received funding from the Mesh Labs, Microsoft, UK.
The authors thank the International Max Planck Research School for Intelligent Systems (IMPRS-IS) for supporting B. Thambiraja.
We would like to thank A. Cseke, T. Alexiadis, T. McConnell, for conducting the user studies.
Further, we also thank B. Kabadayi, P. Kulits, W. Zielonka, M. Diomataris, O.Taheri and P. Mayilvahanan for valuable discussions.

\section{Additional Evaluation}

\medskip
\noindent
\textbf{Additional Holistic-motion synthesis evaluation:}
In addition to comparing against SadTalker~\cite{sadtalker} and Talkshow~\cite{yi2023generating}, we also conduct additional comparisons against DiffPoseTalk~\cite{sun2023diffposetalk}, a closely related concurrent work. 
To this end, we perform a qualitative and perceptual study to evaluate the holistic-motion synthesis and style-similarity.
Specifically, we conducted multiple individual user studies in the form of A/B tests (see \Cref{sec:perceptual_study}).
As we see from~\Cref{tab:perceptual_study_v1}, our method consistently outperforms all baselines in generating natural, realistic holistic motion with highly accurate lip synchronization. 
In the style-similarity studies presented in~\Cref{tab:perceptual_study_v2}, our method matches the performance of Imitator~\cite{imitator} and significantly surpasses DiffPoseTalk~\cite{sun2023diffposetalk}.
Notably, unlike our approach and DiffPoseTalk~\cite{sun2023diffposetalk}, Imitator~\cite{imitator} employs a deterministic regression method that lacks the ability to produce diverse samples or offer editing capabilities.
Additionally, we performed a motion editing experiment using DiffPoseTalk~\cite{sun2023diffposetalk} which is shown in the suppl. video.

\begin{table}[h]
     \resizebox{\columnwidth}{!}{
    \begin{tabular}{cl|cc} \toprule
        & &  \multicolumn{2}{c}{Holistic synthesis} \\ 
        \cmidrule(r){3-4} 
       &\textbf{Method} & \textbf{Face Motion} (\%) &  \textbf{Head motion} (\%) \\ \midrule
        1  & Ours vs SadTalker~\cite{sadtalker}               & $88.13$ & $86.43$ \\
        2  & Ours vs TalkShow~\cite{yi2023generating}         & $90.77$ & $87.96$ \\
        3  & Ours vs DiffPoseTalk*~\cite{sun2023diffposetalk}  & $85.33$ & $90.66$ \\
        \bottomrule
    \end{tabular}
    }
    \captionof{table}{
    \textit{User studies evaluating the naturalness of the facial and head motion.}
    A total of $25$ individuals participated in each A/B user-study.
    * : represents concurrent work.
    }
    \label{tab:perceptual_study_v1}
\end{table}
\begin{table}[h]
    \begin{tabular}{cl|c} \toprule
       &\textbf{Method} & \textbf{Style-similarity} (\%) \\ \midrule
        1  & Ours vs Imitator~\cite{imitator}               & $55.64$  \\
        2  & Ours vs DiffPoseTalk*~\cite{sun2023diffposetalk}  & $89.33$ \\
        \bottomrule
    \end{tabular}
    \captionof{table}{
    \textit{User studies evaluating the style-similarity.}
    A total of $25$ individuals participated in each A/B user-study.
    * : represents concurrent work. 
    }
    \label{tab:perceptual_study_v2}
\end{table}

\medskip
\noindent
\textbf{Impact of Guidance-Scale on facial motion synthesis:}
We investigate the impact of the classifier-free-guidance scale $s$~\cite{ho2022classifierfree} using the 'Lip-sync' and $Div^L$ metrics on the non-personalized facial motion synthesis task.
Lower guidance values yield animations with significantly more diverse motion but inferior lip-sync quality. Conversely, higher guidance values result in high-quality animation with reduced diversity. 
We observe a similar trend in our perceptual evaluation.
We find that the guidance scale $s$ is an effective tool to increase synthesis diversity beyond all baselines with only a small loss of lip-sync accuracy for $0.3\leq s \leq 1.0$.
        
\begin{figure}[h]
 \centering
        \resizebox{\columnwidth}{!}{
        \input{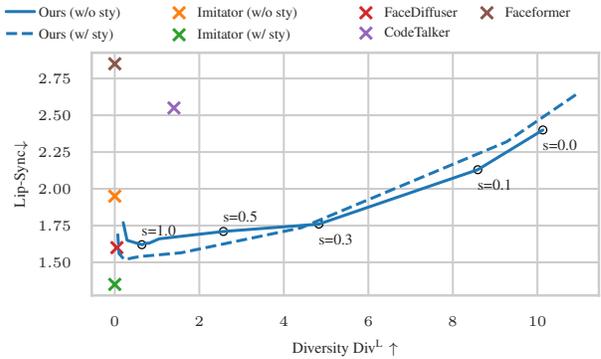}
        }
        \caption{
        Evaluation of the impact of the classifier-free-guidance $s$ on the facial motion synthesis task.
        }
        \label{fig:abl_div_vs_lip_sync}
\end{figure}

\medskip
\noindent
\textbf{Impact of Noise:}
We conducted a noise sensitivity experiment similar to ~\cite{voca,imitator, sun2023diffposetalk}, where we added white noise to the input audio with a negative gain of 36db (low), 24db (medium), and 12db (high).
As reported in~\Cref{tab:noise}, our method is robust to low(36db) and medium(24db) noise levels, produces facial motion with comparable quality.
Please refer to the suppl. video for the qualitative results.
\begin{table}[h]
    \centering
        \begin{tabular}{cl|cc} \toprule
             &\textbf{Method} & $\mathbf{Div^{L}}$ $\uparrow$ & \textbf{Lip-Sync} $\downarrow$ \\ \midrule
                1 & Ours (high noise)   & $6.41$ & $2.56$    \\
                2 & Ours (med. noise) & $2.54$ & $1.97$ \\
                3 & Ours (low noise)    & $1.85$& $1.78$ \\
            \bottomrule
        \end{tabular}
    \vspace{1em}
    \caption{
    Robustness to noise study with low, medium and high noise levels on the VOCAset~\cite{voca}.
    }
    \label{tab:noise}
\end{table}

\begin{figure}[h]
    \centering
    \includegraphics[width=\columnwidth]{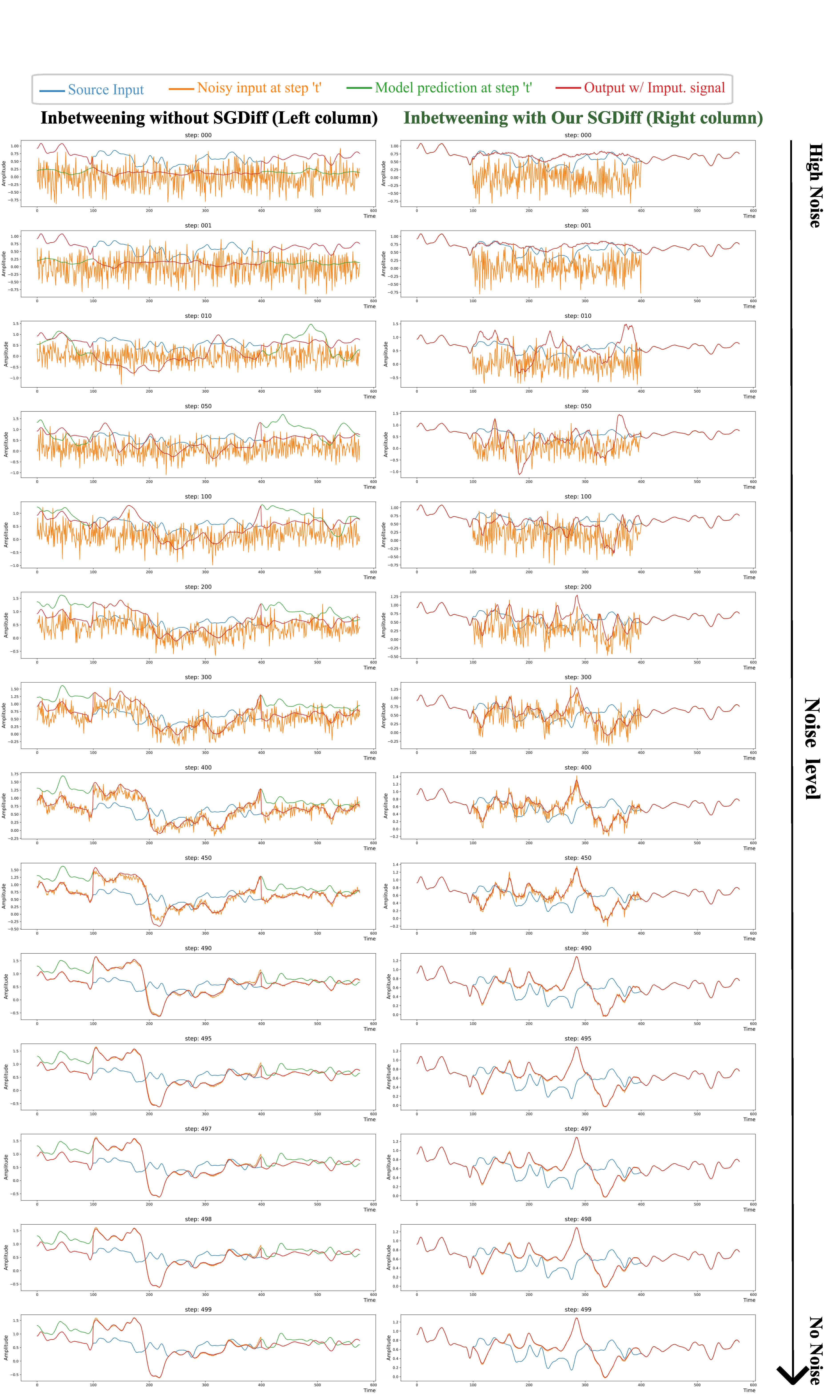}
    \caption{
    Illustration of a conditional head motion inbetweening of a sequence across various diffusion steps \mbox{$t$} with and without Sparsely-guided diffusion. 
    }
    \label{fig:ip_evolution}
\end{figure}

\medskip
\noindent
\textbf{Why standard diffusion-based head motion-editing fails?}
First, we analyze the standard diffusion based head motion editing and show why it fails.
Then, we demonstrate how using our proposed Sparsely-Guided diffusion effectively addresses the issue and enables smoother head-motion editing.
An illustration of a conditional inbetweening of a sequence across various diffusion steps $t$ is shown in the~\Cref{fig:ip_evolution}.
Looking at the left column of the figure, it is clear that the standard diffusion mainly focuses on generating a valid sample from the distribution based on the audio condition. As a result, it starts to ignore the imputation signal in the low noise regime, focusing instead on refining the sequence to produce an improved sample from the distribution.
This approach results in jittery transitions when the imputation signal is replaced to generate the final inbetweened sample at the end of the sampling process.
Throughout its training, the diffusion model was only trained to generate a valid sample from the distribution based on the audio condition, not to align with the imputation signal.
Observing this, we introduced a sparsely-guided diffusion to incorporate guidance signals during training. 
This adjustment ensures that the diffusion model aligns with the imputation signal while still producing a sample from the distribution.

\medskip
\noindent
\textbf{Is it possible to unconditionally synthesize and edit motion?}
\begin{figure}[h!]
    \centering
    \includegraphics[width=0.9\columnwidth]{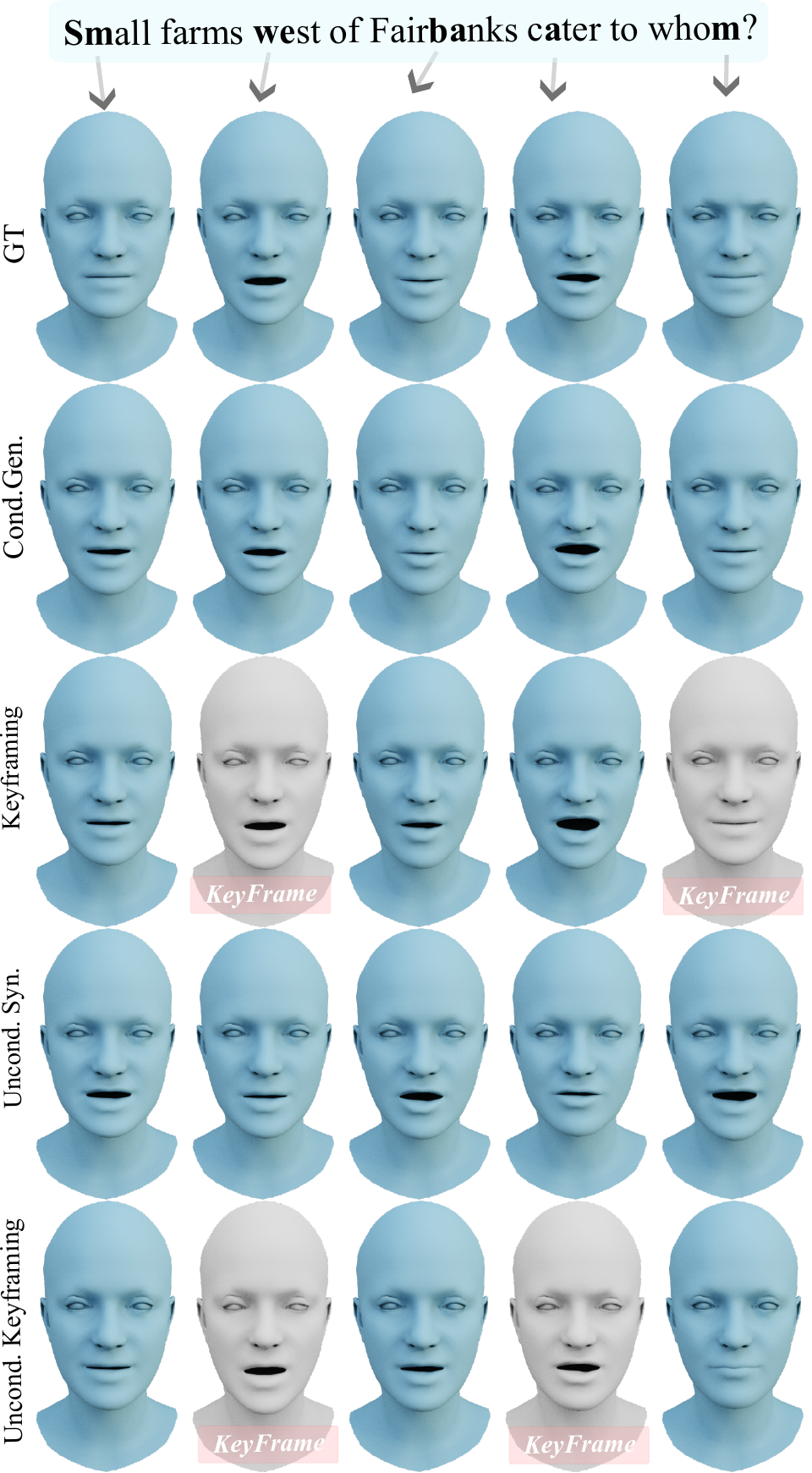}
      \caption{
      Qualitative illustration of facial motion inbetweening using our conditional (row 3) and unconditional model (row 5). 
      }
      \label{fig:ablation_main}
\end{figure}
While unconditional motion synthesis has been extensively applied in the motion synthesis domain~\cite{tevet2023human, raab2022modi}, to the best of our knowledge, its application in 3D facial animation synthesis remains widely unexplored.
The significance of an unconstrained facial motion synthesis method cannot be overstated. It holds substantial potential for various applications, such as animating background characters in movies and games.
Additionally, it enables targeted editing of specific facial elements—such as eye blinks and eyebrow motions—since these non-verbal facial expressions often exhibit weak or no correlation with audio features.
Moreover, an unconditional model serves as a valuable motion prior for various downstream tasks, extending its utility beyond synthesis and editing applications. 
Our demonstration of unconditional synthesis and editing are showcased in~\Cref{fig:ablation_main}, underscoring the potential and versatility of such unconstrained models for 3D facial animation synthesis.
In~\Cref{fig:ablation_main} (rows 2 and 3), we showcase a sequence synthesized conditionally and subsequently refined using keyframes.
In~\Cref{fig:ablation_main} (row 4), we present our unconditional synthesis results. As observed from the results, our model can unconditionally synthesize plausible facial motion.
Additionally, in~\Cref{fig:ablation_main} (row 5), we see that our method can unconditionally inbetween facial animation while preserving the speaking style of the target actor.
This progression illustrate our model's capabilities: from conditional synthesis and keyframe-based editing to unconditional synthesis and editing, while preserving the target actor's speaking style.
Please refer to the supplemental video for the study's results in motion.

\medskip
\noindent
\textbf{Impact of sparse guidance on Facial 3D animation synthesis:}
In this section, we evaluate the impact of using sparsely-guided diffusion (SGDiff) on 3D facial animation synthesis.
To this end, we first trained the face motion generator with SGDiff, then personalized it with and without SGDiff using the VOCAset~\cite{voca} test subject `0024' and report the metrics in \Cref{tab:gmd_3d_facial_animation_synthesis}.
While the introduction of sparse guidance improves diversity, it also reduces the lip-sync accuracy. 
The combination of window-based training and replacement of the input signal within the window acts as data augmentation, increasing the diversity and complexity of the training distribution. As a result, this leads to under-fitting, producing more generalized motion with high-diversity, but poorer lip-sync.
As demonstrated in the main paper, style personalization is critical for 3D facial motion synthesis. Therefore, in our experiments, we employ a personalized facial motion model without sparse-guidance.
\begin{table}[h]
    \centering
    \resizebox{0.8\columnwidth}{!}{%
        \begin{tabular}{cl|cc} \toprule
             &\textbf{Method} & $\mathbf{Div^{L}}$ $\uparrow$ & \textbf{Lip-Sync} $\downarrow$ \\ \midrule
                1 & Ours (without SGDiff) & $1.35$ & $\mathbf{1.4}$  \\ 
                2 & Ours (with SGDiff) & $\mathbf{2.19}$ & $2.44$   \\
            \bottomrule
        \end{tabular}
    }
    \caption{   
    Impact of Sparsely-Guided diffusion on the 3D Facial motion synthesis. Introducing guidance during diffusion improves the diversity and reduces the lip-sync.
    }
    \label{tab:gmd_3d_facial_animation_synthesis}
\end{table}

\medskip
\noindent
\textbf{Impact of Personalization on Head-motion synthesis:}
We evaluate the impact of personalization on head-motion synthesis by personalizing the head-motion using the subjects in the HDTF test-set.
The metric of this evaluation is reported in \Cref{tab:personalization_head_motion}.
From this experiment, we observe that personalization of the head motion often overfits to the speech context and cadence, resulting in high beat alignment and low diversity metrics.
Since, one of our primary goal of this work is to enable diverse motion synthesis, we opted for a non-personalized head-motion model.
This approach offers  better diversity and more flexible editing capabilities.
\begin{table}[h]
    \centering
    \resizebox{0.9\columnwidth}{!}{%
        \begin{tabular}{cl|cc} \toprule
                 &\textbf{Method} & $\mathbf{BA}$ $\uparrow$  & $\mathbf{Div^{H}}$ $\uparrow$  \\ \midrule
                1 & Ours w/o. Personalization & $0.338$ &  $\mathbf{0.007}$\\
                2 & Ours w. Personalization & $\mathbf{0.673}$ &  $0.002$\\
            \bottomrule
        \end{tabular}
    }
    \caption{   
    Impact of personalization on the head-motion synthesis.
    }
    \label{tab:personalization_head_motion}
\end{table}

\begin{figure}[h]
    \centering
   \resizebox{0.8\columnwidth}{!}{
    \input{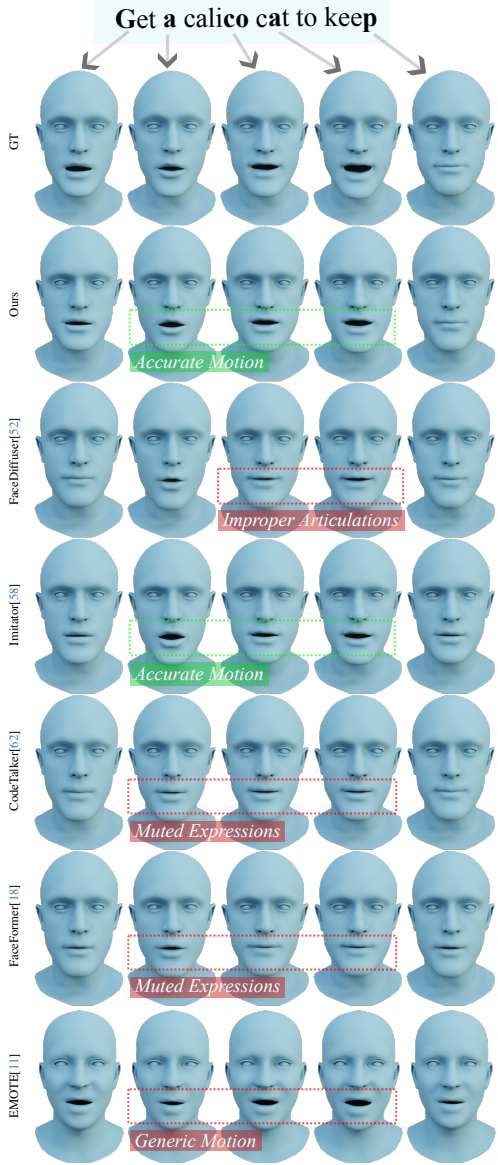}
    }
  \caption{
  Qualitative comparison of the facial motion synthesis model. 
    }
  \label{fig:baseline_comp}
\end{figure}

\medskip
\noindent
\textbf{Qualitative comparison against 3D facial motion synthesis methods:}
As mentioned in the main paper, we evaluate our method against the state-of-the-art methods  VOCA~\cite{voca}, Faceformer~\cite{fan2022faceformer}, CodeTalker~\cite{xing2023codetalker}, EMOTE~\cite{EMOTE}, FaceDiffuser~\cite{FaceDiffuser_Stan_MIG2023} and Imitator~\cite{imitator} on facial motion synthesis task.
A qualitative comparison to the facial motion synthesis baselines on a test sequence from the VOCAset is shown in \Cref{fig:baseline_comp}, where our method produces expressive facial animations that match the speaking style of the target subjects. 

\medskip
\noindent
\textbf{Quantitative evaluation on the BIWI Dataset:}
The main focus of our work is to enable diverse synthesis with precise control.
This made both BIWI~\cite{eth_biwi_00760} and BEAT~\cite{liu2022beat} incompatible for our study, as they are in different model spaces compared to the existing face trackers like~\cite{EMOCA_CVPR_2021, mica, filntisis2022visual}.
This is a key necessity for personalization and subsequently face motion editing.
More details about this is discussed in \Cref{data:discussion}.
Nevertheless, for completeness, we conduct a quantitative evaluation our 3D facial motion synthesis against the state-of-the-art methods trained on BIWI~\cite{eth_biwi_00760}, in addition to the quantitative comparison presented on VOCA~\cite{voca} in the main paper.
To this end, we adopt the dataset setup used by~\cite{fan2022faceformer, xing2023codetalker, FaceDiffuser_Stan_MIG2023} and only use the emotional sequence subset.
Specifically, the data is split into a training set (BIWI-Train) containing 192 sentences and a validation set (BIWI-Val) with 24 sentences from 6 training subjects. 
There are two test sets: BIWI-Test-A, containing 24 sentences from seen subjects, and BIWI-Test-B, containing 32 sentences from 8 unseen subjects.
For this experiment, we perform the qualitative study on the BIWI-Test-A and report the metric in~\Cref{tab:personalization_head_motion}.
We use the Lip vertex error ($LVE$) metric used by the baselines~\cite{FaceDiffuser_Stan_MIG2023, xing2023codetalker} in this experiment.
From the results, we can observe that our method outperforms the baselines in terms of producing high-quality lip motion.
\begin{table}[h]
    \centering
    \begin{tabular}{cl|cc} \toprule
            &\textbf{Method} & $\mathbf{LVE}$ $\downarrow$  \\ \midrule
            1 & VOCA~\cite{voca}        & $6.55$ \\
            2 & MeshTalk~\cite{meshtalk}    & $5.91$ \\
            3 & FaceFormer~\cite{fan2022faceformer}  & $5.3$ \\
            4 & CodeTalker~\cite{xing2023codetalker}  & $4.79$ \\
            5 & FaceDiffuser~\cite{FaceDiffuser_Stan_MIG2023} & $4.29$ \\
            6 & Ours    & $\mathbf{3.61}$ \\	
            \bottomrule
        \end{tabular}
    \caption{   
    Impact of personalization on the head-motion synthesis.
    }
    \label{tab:personalization_head_motion}
\end{table}

\section{Implementation}
In this section, we provide more details on the diffusion model, dataset, baselines, and metrics. 

\subsection{Preliminaries}
\label{sec:preliminaries}

\paragraph{Denoising Diffusion Probabilistic Models:}
Our method is based on the diffusion framework of Sohl et al.~\cite{sohl2015deep}, where a training sample \(x_0\) gradually transforms into white noise through the addition of Gaussian noise across \(T\) steps. This transformation is mathematically represented as: 
\begin{equation}
    x_t \sim q(x_t|x_{t-1}) = \mathcal{N}(x_t; \sqrt{ 1 - \beta_{t}} x_{t-1},  \beta_{t}I), t=1...T ,
\end{equation}
where $\beta_t$ is following a predefined variance schedule.

Following recent work~\cite{tevet2023human, sun2023diffposetalk}, we train a denoising model $\theta$ that can reverse this noisy diffusion and estimate the original sample \(x_0\) from a noised version \(x_t\), guided by: $ \hat{x}_0 = \theta(x_t, t, C)$.
With \(\theta\) being the neural network and \(C\) representing additional conditions. The reverse diffusion is achieved through:
\[
    q(x_{t-1}|x_t) = \mathcal{N}\left(x_{t-1}; \sqrt{\bar{\alpha}_{t-1}}\theta(x_t, t, C), (1-\bar{\alpha}_{t-1}) I\right),
\]
where \(\alpha_t:=1-\beta_t\) and \(\bar{\alpha}_t:=\prod_{k=1}^t\alpha_k\).

To generate new samples, we start from random noise \(x_T\) and apply iterative denoising until reaching \(t=0\). We introduce diversity in generation using Classifier-Free Guidance (CFG)~\cite{ho2022classifierfree} by combining conditional and unconditional predictions of the network, controlled by a guidance scale \(s\):
\[
    \theta_s(x_t, t, C) := \theta(x_t, t, \emptyset) + s \cdot \left[\theta(x_t, t, C)-\theta(x_t, t, \emptyset)\right],
\]
adjusting \(s\) to balance between diversity and adherence to conditions.
Following \cite{ho2020denoising}, the inverse diffusion process is then given through:
\begin{equation}
    q(x_{t-1}|x_t) = \mathcal{N}\left(x_{t-1}; \sqrt{\bar{\alpha}_{t-1}}\theta(x_t, t, C), (1-\bar{\alpha}_{t-1}) I\right) ,
\end{equation}
with $\alpha_t:=1-\beta_t$ and $\bar{\alpha}_t:=\prod_{k=1}^t\alpha_k$.
For generating new samples, we randomly sample $x_T$ from a Gaussian distribution and iteratively denoise it until $t=0$ is reached.

To add diversity, we employ Classifier-Free Guidance (CFG)~\cite{ho2022classifierfree} and calculate the output as a weighted sum of the conditional and unconditional prediction:
\begin{equation}
    \theta_s(x_t, t, C) := \theta(x_t, t, \emptyset) + s \cdot \left[\theta(x_t, t, C)-\theta(x_t, t, \emptyset)\right] ,
\end{equation}
where $s$ is the guidance scale and $\theta(x_t, t, \emptyset)$ denotes the unconditional prediction in which we set the audio conditions to zero.
Note that while CFG is typically used with a guidance scale $>1$ to enhance alignment with the condition, we set it to values $<1$ (0.5 unless specified otherwise) to increase diversity.

\subsection{Dataset}
\paragraph{VOCA:}%
We train our facial motion model on the VOCAset~\cite{voca} since it provides high-quality, speech-aligned 3D face scan sequences. 
It consists of $12$ actors ($6$ female and $6$ male) with $40$ sequences each with a length of $3$-$5$ seconds, resampled at $30$fps.
Following previous work \cite{imitator}, we use the train/val/test set split of $8,2,2$ actors.
All $40$ sequences of the training actors are used during training. 
However, for the test and validation, only $20$ sequences that do not overlap with the speech scripts of the training sequences are used.
For the style adaption experiment, we split the $40$ sequences of the test actors to $18, 2, 20$ for train/val/test sets. 
The test sequences of the experiments w/ and w/o style adaptation are identical, allowing a direct comparison of the scores in the quantitative comparison in the main paper (Table 2). 

\paragraph{In-the-wild dataset: }%
We evaluate person-specific fine-tuning on in-the-wild video sequences from  Imitator~\cite{imitator}. The provided videos are 2 minutes long which we divide into 60/30/30 seconds for train/val/test respectively.
Similar to Imitator, we employ the MICA tracker~\cite{mica} to extract the face motion tracking for the personalization step. 

\paragraph{HDTF: }%
We train our head-motion generator on the HDTF~\cite{HDTF} dataset.
The High-definition Talking Face Dataset (HDTF) is a large in-the-wild audio-visual dataset for talking face generation.
It consists of about 362 different high-resolution ($720$P or $1080$P) YouTube videos of $15.8$ hours in total. 
Using the download and processing script provided by the authors, we extracted 352 videos with 246 unique subjects.
We additionally crop the video to 30 seconds long and use them for extracting head-poses using the MICA tracker~\cite{mica}, which provides head poses as global axis rotation.
For our experiments, we split the dataset into 300/20/32 sequences for train/val/test accordingly.

\paragraph{Discussion: }
\label{data:discussion}
In this work, we employ the VOCAset, HDTF, and Imitator's in-the-wild dataset to train our method for generating and editing 3D facial animations with head motion. 
The motivation of generating and editing holistic 3D facial animation made both BIWI~\cite{eth_biwi_00760} and BEAT~\cite{liu2022beat} incompatible for our study, both BIWI and BEAT are in different model spaces compared to the existing face trackers like~\cite{EMOCA_CVPR_2021, mica, filntisis2022visual}, which is a key necessity for personalization and subsequently face motion editing.
Such a problem could in theory be addressed by converting the meshes provided in the dataset to our target FLAME model space by optimization-based fitting using pre-defined correspondence between the source and target mesh space.
However, for BIWI the noisy surface reconstructions provided in the dataset and incomplete face models make the fitting challenging and reduces the quality of the fitted meshes further.
Similarly, for BEAT, the dependence on ARKit which produces improper lip-closures and not fully completed face model, reduces the realism of the reconstructed sequences.
As studied in~\cite{imitator}, lip-closures are paramount in conveying realism for the generated sequences.

\subsection{Baselines}
\paragraph{Holistic 3D motion synthesis:}
For TalkShow~\cite{yi2023generating}, we use the pre-trained model provided in their official repository and extract the predicted facial and head motion parameters for our evaluation.
For SadTalker~\cite{sadtalker}, we use the pre-trained model provided in the repository to generate 2D talking face videos and use the MICA tracker~\cite{mica} to the face and head motion.
For DiffPoseTalk~\cite{sun2023diffposetalk}, we use the pre-trained provided in their repository. Similar to our work, DiffPoseTalk~\cite{sun2023diffposetalk} requires 3D reconstructions of target subjects for style-personalizaton. For 3D reconstruction, DiffPoseTalk utilize their own customer-tracker, which, at the time of the submission was not available in their repository. Hence, we substitute their tracker with EMOCAv2~\cite{EMOCA_CVPR_2021}, a publicly available face tracker in our experiments.

\paragraph{Facial Motion Synthesis:}
For VOCA~\cite{voca}, Faceformer~\cite{fan2022faceformer}, Imitator~\cite{imitator} and FaceDiffuser~\cite{FaceDiffuser_Stan_MIG2023}, we use the pre-trained model provided in the official repositories.
For CodeTalker~\cite{xing2023codetalker}, we adapt the official implementation to add the functionality of generating diverse motion. Especially, we re-train the audio-conditioned codebook sampling (stage 02) to randomly sample a code from the top 'm' closest codes instead of always using the closest code.
This process is in spirit close to training the language-based models, where a new diverse text sequence is generated by sampling the 2nd or 3rd closest language token over the token with maximum probability.
By adapting this method, we ensure that CodeTalker can generate diverse samples for a given audio input.
For EMOTE~\cite{EMOTE}, we request the authors to run their method on the VOCAset~\cite{voca} and use it for the qualitative and perceptual user study.

\subsection{Training Details}
\paragraph{Facial Motion Synthesis:}
We train our method using ADAM~\cite{kingma2017adam} with a learning rate of \textit{1e-4} for 140K iterations with a batch size of $64$.
Our diffusion framework is based on the Gaussian diffusion from Nichol \textit{et al.}~\cite{nichol2021improved}, we set the diffusion step to $500$ for our experiments.
During training, we randomly crop the sequences to the length of $30$ frames.
Our lightweight architecture enables us to train our model on a single Nvidia Quadro P6000 $32GB$ within $30$ hours.
The lightweight architecture is also critical for person-specific style adaptation with a short reference video.
For person-specific speaking style, we use the same training setup as from the generalized setting, except that we only train it for $30K$ iterations.
For evaluating the best checkpoint, we fix the guidance scale $s=0.99$ and evaluate all the saved checkpoints on the validation set.
Further, we fix the best checkpoint and vary the guidance scale from \textit{s=0, 0.1 ... to 1.0} with an increment of $0.1$ and find the best guidance factor.
From our experiment, we found the guidance scale of $0.5$ balances the lip-synchronization and diversity and provides the best results.

\paragraph{Head Motion Synthesis}
Similar to the Facial motion synthesis pipeline, we train our method using ADAM~\cite{kingma2017adam} with a learning rate of \textit{1e-4} for 100K iterations with a batch size of $64$.
During training the sequences are randomly cropped to $300$ frames long.
For our inbetweening and keyframing-based Guided motion model training, we randomly sample a mask of arbitrary length for imputation signal, using which the noisy input is replaced with the ground truth imputation signal.

\subsection{Inference}
Our method takes $3.15$ sec to produce $1$ sec ($30$ frame) of facial motion and $1.04$ sec to produce $1$ sec ($30$ frame) of head motion on a single Nvidia GeForce RTX 3090 $24GB$, compared to $5.78$ sec for the concurrent method FaceDiffuser~\cite{FaceDiffuser_Stan_MIG2023}. 
In total, our method takes $4.19$ sec to produce $1$ sec ($30$ frame) of holistic 3D facial animation, compared to $6.78$ sec for TalkSHOW~\cite{yi2023generating}.
\subsection{Metrics}
\paragraph{Lip-Sync} measures the lip synchronization using Dynamic Time Warping to compute the temporal similarity~\cite{imitator}.

\paragraph{Diversity metric} introduced by Ren et al.~\cite{ren2023diffusion} measures the diversity of 3D motions for the same text input.
We employ this metric and propose $Div^L$ and $Div^H$ to measure the diversity of lip motion and head motions generated from the same audio.
Given a set of generated 3D facial or head motions with $N$ sequences generated from the same audio condition.
The diversity can be formalized as:
\begin{equation}
Diversity =  \frac{1}{L} \sum_{i=1}^{N-1} \sum_{j=i+1}^{N} \left\lVert m_{i} - m_{j} \right\rVert_2
\end{equation}
Where $m_{i}$ represents the $i$-th motion and $L$ is the total number of possible combinations in the generated motion set.

\paragraph{Beat alignment (BA)}: Similar to DiffPoseTalk~\cite{sun2023diffposetalk}, we employ a modified beat alignment \textit{BA} to measure the synchronization of the head movement beats between the predicted and ground truth motion, where we calculate the average temporal distance between beat in predicted head movement its closest ground truth beat as the Beat Align Score.
\begin{equation}
    \text{Beat Align Score} = \frac{1}{|\mathbf{B}_g|} \sum_{t_g \in \mathbf{B}_g} \exp \left( -\frac{\min_{t_p \in \mathbf{B}_p} \lVert t^p - t^g \rVert_2^2}{2\sigma^2} \right),
\end{equation}
Where $\mathbf{B}_g$ and $\mathbf{B}_p$ record the time of the beats in the ground truth and predicted head motion respectively, while $\sigma$ is the normalized parameter which is set to be $3$ in our experiment.

\paragraph{Discussion}
The $L2$-based vertex error metrics employed in previous studies~\cite{imitator, fan2022faceformer, xing2023codetalker} are not apt for our task due to its preference for solutions that are close to the mean of the dataset, which penalizes the diversity present in our predictions.
\subsection{Perceptual Study}
\label{sec:perceptual_study}
\begin{figure}[ht!]
    \centering
    \includegraphics[width=\linewidth]{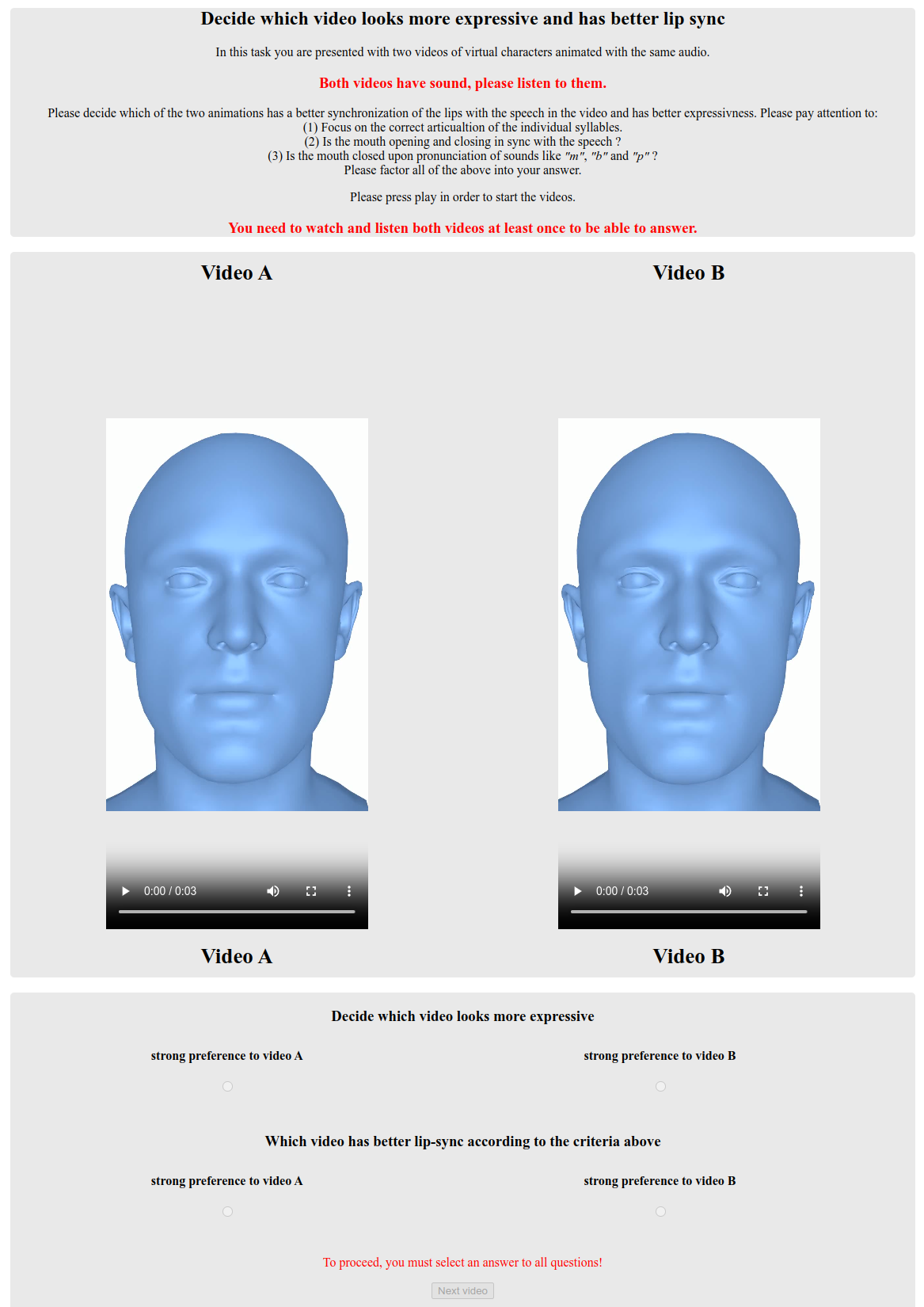}
    \caption{Example of the interface employed for our user-study. }
    \label{fig:user_study}
\end{figure}

We conducted A/B user studies to assess our method's perceptual performance.
First, we conducted a study to evaluate the holistic motion synthesis based on the naturalness of the facial and head motion to the input audio.
For this, we sampled $10$ sequences from the test set of the HDTF and $10$ external audio from YouTube and synthesized holistic 3D facial motion using our method and the baselines~\cite{sadtalker, yi2023generating, sun2023diffposetalk} resulting in a total of $60$ A/B comparisons including ground truth.
For extracting the ground truth for the YouTube sequences, similar to the HDTF dataset processing we utilize the MICA tracker~\cite{mica} to extract the facial and head motion.
Through Amazon Mechanical Turk(AMT) and Google forms, we divided the A/B comparisons into $3$ HITs~(Human Intelligence Task), each with $25$ individual assignments. For each HIT, users were instructed to select their preference for a method based naturalness of the head and facial motion with synchronization.
Second for facial motion synthesis, we sample 20 sequences combined from the VOCAset test set and the in-the-wild sequences from Imitator, resulting in 100 A/B comparisons across five baselines.
On Amazon Mechanical Turk(AMT), we divided the A/B comparisons into $5$ HITs~(Human Intelligence Task), each with $25$ individual assignments. For each HIT, users select their preference for a method based on expressiveness and lip-synchronization.
Finally, we evaluated the speaking style preservation of our personalized model in comparison to Imitator. 
To this end, the AMT users rated the similarity based on a reference video and the synthesized videos of the VOCA test set.
~\Cref{fig:user_study} illustrates an example interface in our user-study.

\section{Ethical Impact}
\label{sec:broader_impact}
We introduce a method for realistic facial animation synthesis and editing that matches the speaking style of any given target actor.
These animations hold promise for driving virtual avatars in AR or VR settings, especially, in immersive communication technologies.
Yet, it is essential to acknowledge the potential pitfalls of such advancements, notably in the realm of 'DeepFakes.' 
By employing voice cloning techniques, our method can generate 3D facial animations that drive digital avatar methods like~\cite{Gafni_2021_CVPR, Zielonka2022InstantVH, grassal2022neural, kabadayi24ganavatar, bharadwaj2023flare}, which could be abused for identity theft, cyberbullying, and various criminal activities.
Advocating for transparent research practices, we strive to illuminate the risks associated with technology misuse. 
Sharing our implementation aims to foster research in digital multimedia forensics, particularly in developing synthesis methods crucial for training data utilized in spotting forgeries~\cite{roessler2019faceforensics++}.

{
    \small
    \bibliographystyle{ieeenat_fullname}
    \bibliography{main}
}

\end{document}